\documentclass[12pt,preprint]{aastex61}
\usepackage{graphicx}
\usepackage{epstopdf}

\usepackage{subfigure}

\usepackage{bm}

\usepackage{natbib}
\usepackage{color}

\shorttitle{A Brown-Dwarf Binary at the Deuterium Fusion Limit}
\shortauthors{Albrow}

\begin{document}

\title{OGLE-2016-BLG-1266: A Probable Brown-Dwarf/Planet Binary at the Deuterium Fusion Limit}

\author{M. D. Albrow}
\affiliation{School of Physical and Chemical Sciences, University of Canterbury, Private Bag 4800, Christchurch, New Zealand}

\author{J. C. Yee}
\affiliation{Harvard-Smithsonian Center for Astrophysics, 60 Garden St., Cambridge, MA 02138, USA}

\author{A. Udalski}
\affiliation{Warsaw University Observatory, Al. Ujazdowskie 4, 00-478 Warszawa, Poland}

\nocollaboration

%
%

\author{S. Calchi Novati}
\affiliation{IPAC, Mail Code 100-22, Caltech, 1200 E. California Blvd., Pasadena, CA 91125, USA}

\author{S. Carey}
\affiliation{Spitzer, Science Center, MS 220-6, California Institute of Technology,Pasadena, CA, USA}

\author{C. B. Henderson}
\affiliation{Jet Propulsion Laboratory, California Institute of Technology, 4800 Oak Grove Drive, Pasadena, CA 91109, USA}

\author{C. Beichman}
\affiliation{NASA Exoplanet Science Institute, California Institute of Technology, Pasadena, CA 91125, USA}

\author{G. Bryden}
\affiliation{Jet Propulsion Laboratory, California Institute of Technology, 4800 Oak Grove Drive, Pasadena, CA 91109, USA}

\author{B. S. Gaudi}
\affiliation{Department of Astronomy, Ohio State University, 140 W.18th Ave., Columbus, OH 43210, USA}

\author{Y. Shvartzvald}
\affiliation{Jet Propulsion Laboratory, California Institute of Technology, 4800 Oak Grove Drive, Pasadena, CA 91109, USA}
\affiliation{NASA Postdoctoral Program Fellow}

\collaboration{({\it Spitzer} team)}

%
%

\author{M. K. Szyma\'{n}ski}
\affiliation{Warsaw University Observatory, Al. Ujazdowskie 4, 00-478 Warszawa, Poland}

\author{P. Mr\'{o}z}
\affiliation{Warsaw University Observatory, Al. Ujazdowskie 4, 00-478 Warszawa, Poland}

\author{J. Skowron}
\affiliation{Warsaw University Observatory, Al. Ujazdowskie 4, 00-478 Warszawa, Poland}

\author{R. Poleski}
\affiliation{Warsaw University Observatory, Al. Ujazdowskie 4, 00-478 Warszawa, Poland}
\affiliation{Department of Astronomy, Ohio State University, 140 W.18th Ave., Columbus, OH 43210, USA}

\author{I. Soszy\'{n}ski}
\affiliation{Warsaw University Observatory, Al. Ujazdowskie 4, 00-478 Warszawa, Poland}

\author{S. Koz{\l}owski}
\affiliation{Warsaw University Observatory, Al. Ujazdowskie 4, 00-478 Warszawa, Poland}

\author{P. Pietrukowicz}
\affiliation{Warsaw University Observatory, Al. Ujazdowskie 4, 00-478 Warszawa, Poland}

\author{K. Ulaczyk}
\affiliation{Warsaw University Observatory, Al. Ujazdowskie 4, 00-478 Warszawa, Poland}

\author{M. Pawlak}
\affiliation{Warsaw University Observatory, Al. Ujazdowskie 4, 00-478 Warszawa, Poland}
\collaboration{(OGLE Collaboration)}

%
%
\author{S.-J. Chung}
\affiliation{Korea Astronomy and Space Science Institute, Daejon 34055, Korea}
\affiliation{Korea University of Science and Technology, 217 Gajeong-ro, Yuseong-gu, Daejeon, 34113, Korea}

\author{A. Gould}
\affiliation{Department of Astronomy, Ohio State University, 140 W.18th Ave., Columbus, OH 43210, USA}
\affiliation{Korea Astronomy and Space Science Institute, Daejon 34055, Korea}
\affiliation{Max-Planck-Institute for Astronomy, K\"{o}nigstuhl 17, 69117 Heidelberg, Germany}

\author{C. Han}
\affiliation{Department of Physics, Chungbuk National University, Cheongju 28644, Republic of Korea}

\author{K.-H. Hwang}
\affiliation{Korea Astronomy and Space Science Institute, Daejon 34055, Korea}

\author{Y. K. Jung}
\affiliation{Harvard-Smithsonian Center for Astrophysics, 60 Garden St., Cambridge, MA 02138, USA}

\author{Y.-H. Ryu}
\affiliation{Korea Astronomy and Space Science Institute, Daejon 34055, Korea}

\author{I.-G. Shin}
\affiliation{Harvard-Smithsonian Center for Astrophysics, 60 Garden St., Cambridge, MA 02138, USA}

\author{W. Zhu}
\affiliation{Canadian Institute for Theoretical Astrophysics, 60 St. George Street, University of Toronto, Toronto, ON M5S 3H8,
Canada}

\author{S.-M. Cha}
\affiliation{Korea Astronomy and Space Science Institute, Daejon 34055, Korea}
\affiliation{School of Space Research, Kyung Hee University,Yongin, Kyeonggi 17104, Korea}

\author{D.-J. Kim}
\affiliation{Korea Astronomy and Space Science Institute, Daejon 34055, Korea}

\author{H.-W. Kim}
\affiliation{Korea Astronomy and Space Science Institute, Daejon 34055, Korea}

\author{S.-L. Kim}
\affiliation{Korea Astronomy and Space Science Institute, Daejon 34055, Korea}
\affiliation{Korea University of Science and Technology, 217 Gajeong-ro, Yuseong-gu, Daejeon, 34113, Korea}

\author{C.-U. Lee}
\affiliation{Korea Astronomy and Space Science Institute, Daejon 34055, Korea}
\affiliation{Korea University of Science and Technology, 217 Gajeong-ro, Yuseong-gu, Daejeon, 34113, Korea}

\author{D.-J. Lee}
\affiliation{Korea Astronomy and Space Science Institute, Daejon 34055, Korea}
\affiliation{Korea University of Science and Technology, 217 Gajeong-ro, Yuseong-gu, Daejeon, 34113, Korea}

\author{Y. Lee}
\affiliation{Korea Astronomy and Space Science Institute, Daejon 34055, Korea}
\affiliation{School of Space Research, Kyung Hee University,Yongin, Kyeonggi 17104, Korea}

\author{B.-G. Park}
\affiliation{Korea Astronomy and Space Science Institute, Daejon 34055, Korea}
\affiliation{Korea University of Science and Technology, 217 Gajeong-ro, Yuseong-gu, Daejeon, 34113, Korea}

\author{R. W. Pogge}
\affiliation{Department of Astronomy, Ohio State University, 140 W.18th Ave., Columbus, OH 43210, USA}

\collaboration{(KMTNet Collaboration)}

\correspondingauthor{M. D. Albrow}
\email{Michael.Albrow@canterbury.ac.nz}

\begin{abstract}
We report the discovery, via the microlensing method, of a new very-low-mass binary system.
By combining measurements from Earth and from the {\it Spitzer} telescope in Earth-trailing orbit, 
we are able to measure the microlensing parallax of the event, and find that the lens likely consists 
of an $(12.0 \pm 0.6) M_{\rm J}$ + $(15.7 \pm 1.5) M_{\rm J}$ super-Jupiter / brown-dwarf pair. The binary is located
at a distance of $(3.08 \pm 0.18)$ kpc in the Galactic Plane, and the components have a projected separation of
 $(0.43 \pm 0.03)$ AU. 
 
Two alternative solutions with much lower likelihoods are also discussed,  an 8-  and 6-$M_{\rm J}$ model
and a 90-  and 70-$M_{\rm J}$ model. 
Although disfavored at the 3-$\sigma$ and 5-$\sigma$ levels, these alternatives cannot be rejected
entirely. We show how the more-massive of these models could be tested with future direct imaging.

\end{abstract}

\keywords{}

\section{Introduction}

The growing number of detections of super-Jupiter-mass objects, both isolated and in orbit around objects of higher mass,
raises challenges of interpretation and classification.

Formal definitions of what constitutes a ``planet'' tend to be based
on the mass or interior physics of the object. 
The IAU Working Group on Extrasolar Planets (which existed until 2006) considered the deuterium fusion limit 
($\sim 13 M_{\rm J}$ for solar metallicity) to be the dividing line between planets and brown dwarfs 
for objects that orbit stars. They also considered substellar objects with masses above 
the deuterium fusion limit to always be brown dwarfs. 
The NASA Exoplanet Archive adopts a looser definition for inclusion in their planet tables, namely that the inclusion of
an object as a planet is made provided that its mass is less that 30 $M_{\rm J}$ and it is associated with a host
star\footnote{Confusingly,  the Archive violates its own policy by including objects with brown-dwarf hosts. 
https://exoplanetarchive.ipac.caltech.edu/docs/exoplanet\_criteria.html}. 

 On the other hand, the logical definition for what constitutes a
  ``planet'' would be based on formation mechanism, i.e., whether the
  object formed in a disk or through direct collapse of the gas
  cloud. This might suggest a distinction between super-Jupiter-mass
  objects that orbit stars and those that orbit hosts of comparable
  mass (i.e., very low mass, brown dwarf-brown dwarf binaries), and
  raises questions about how to classify those without hosts.  In
  fact, the observational community tends to make a distinction
  between super-Jupiters orbiting stars and those orbiting brown
  dwarfs.
\citet{Best2017} refer to 2MASS J11193254-1137466 (a member of the TW Hydrae Association) 
as a pair of 3.7 $M_{\rm J}$  brown dwarfs, and suggest that the system is 
a product of normal star formation processes.  In contrast, 
\citet{Lovis2007} refer to the 10.6 $M_{\rm J}$ object orbiting the 2.4 $M_\sun$ star TYC 5409-2156-1 as a planet, and 
argue that an abrupt transition between planets and brown dwarfs has little meaning
if both categories of objects are formed by the same physical
process. Likewise \citet{Carson2013} argue that a planetary classification rather than brown dwarf is 
appropriate for a 12.8 $M_{\rm J}$  body orbiting the  2.5 $M_\sun$ host star, $\kappa$ And.

Formal definitions do not capture these nuances. The IAU makes a
  specific distinction for isolated objects located in young star
  clusters: below the deuterium-burning limit, they are classified as
  ``sub-brown dwarfs'' \citep{Boss2007}. However, the classification
  of an object at or below the deuterium fusion limit that is
  gravitationally bound to another sub-stellar object is not currently
  defined by the IAU. Neither is the case of an isolated object of
  that mass located outside a young cluster.

Precise definitions are complicated by the fact that without
  observing the actual formation of the objects, it is impossible to
  say what mechanism led to their formation and where the boundary
  should be. For example, \citet{Mordasini2009}  show that it is
  theoretically feasible to grow super-Jupiters by core accretion in a
  proto-planetary disk up to at least $30 \, M_{\rm J}$. At the same time,
  \citet{Schlaufman2018} has recently suggested that any companions to
  solar-type stars with mass $> 10 \, M_{\rm J}$ should not be considered
  planets, i.e., could not have formed by core accretion. However,
  since the Schlaufman study was based solely on transiting
  (i.e., short period) objects, it is unclear whether or not this
  result truly reflects something about formation rather than the
  subsequent migration of the objects. Defining the boundary between
  ``planets'' and ``brown dwarf companions'' is further complicated by the
  question of whether or not gravitational instability of a disk
  should be considered to form planets or brown dwarfs. Certainly,
  though, the choice of the deuterium fusion limit as the planet /
  brown-dwarf boundary is arbitrary and confusing \citep{Baraffe2008}.

  It is unclear at this time whether brown-dwarf/super-Jupiter
  binaries belong to the population of objects that formed like binary
  stars from the collapse of molecular clouds, or if some other mechanism, 
  such as ejection from a higher-multiplicity system is
  responsible. Hydrodynamic simulations of the collapse of a large
  star-forming molecular cloud by \citet{Bate2012} resulted in $\sim$ 450 stars
  and $\sim$ 800 brown dwarfs. Of these, some brown dwarfs with masses
  below $30 \, M_{\rm J}$ were formed but no binaries with primary masses below
  $70 \, M_{\rm J}$.

  Understanding the differences in super-Jupiter-mass objects as a
  function of their host mass requires the discovery of more such
  objects, especially those with very-low-mass (brown dwarf) hosts.
  In addition to the \citet{Best2017} binary, a handful of very-low-mass
  binaries have been detected by photometric methods in young open
  clusters and star forming regions \citep{Luhman2013}. However, mass
  estimates for these objects rely on theoretical models of their
  evolution. The uncertainties are large, and the results are strongly
  dependent on the assumed age of the systems.

  Microlensing offers an entirely different avenue for probing the
  population of very-low-mass binaries. In recent years three very-low
  mass binaries have been detected through the channel of
  gravitational microlensing \citep{Choi2013, Han2017b}. In
  contrast to the photometric detections, microlensing binaries can
  have direct and reliable mass estimates, independent of brown-dwarf
  evolutionary theory. Furthermore, these objects are located at large
  distances in the Galactic Disk and are outside of young,
  star-forming clusters.

  In this paper we report the detection of a fourth very-low mass
  binary system by microlensing. This new system is composed of a
  15.7-$M_{\rm J}$ brown dwarf plus a companion just below the deuterium fusion
  limit.

\section{Gravitational Microlensing}

Gravitational microlensing is an effect for which the brightness of a distant star (the source) is magnified due to the bending of light by 
the gravity of a nearer object (the lens). Typically, $> 2000$ microlensing events are detected each year in the direction of
the Galactic Bulge by the 
OGLE\footnote{http://ogle.astrouw.edu.pl/ogle4/ews/ews.html}, 
MOA\footnote{https://www.massey.ac.nz/$\sim$iabond/moa/alerts/} and 
KMTNet\footnote{http://kmtnet.kasi.re.kr/kmtnet-eng/} surveys.

The characteristic angular scale for microlensing is the Einstein radius, 
\begin{equation}
\theta_{\rm E} = \sqrt{ \frac{4GM}{c^2} \left( \frac{1}{D_L } - \frac{1}{D_S} \right) } = \sqrt{ \kappa M \pi_{\rm rel}}, 
\end{equation}
where $M$ is the total mass of the lens system,  $D_L$, $D_S$ are the distances from Earth to the lens and source, 
$\pi_{\rm rel} = \left( \frac{1}{D_L } - \frac{1}{D_S} \right) {\rm AU}$ is the lens-source relative parallax,
and $\kappa  = \frac{4 G}{ c^2 {\rm AU}} = 8.14 \, {\rm mas}/M_\sun$.

The magnification, $A(t)$,
of a standard binary microlensing event can be described by seven parameters in the lens frame. These represent the angular separation of the lens
components ($s$), their mass ratio ($q$), the angular source radius in units of $\theta_{\rm E}$ ($\rho$), the angle of the source trajectory from the lens axis ($\alpha$),
the angular distance of closest approach of the source to the lens center of mass in units of $\theta_{\rm E}$ ($u_{\rm 0}$), the time of closest approach ($t_{\rm 0}$) and 
the Einstein radius crossing time ($t_{\rm E}$). Two additional linear parameters, the source and blend flux $f_{\rm S}$ and $f_{\rm B}$,  are required for each data set
to map the magnification onto the observed flux $f(t)$,  i.e.,
\begin{equation}
f(t) = f_{\rm S} A(t) + f_{\rm B}.
\label{eqn:flux}
\end{equation}

If the source angular radius, $\theta_*$, can be measured independently (usually from its color and an assumption that
it lies behind the same column of dust as the Galactic Bulge),
then the angular Einstein radius $\theta_{\rm E} = \theta_* / \rho$ can be determined.

Additionally, if the microlensing event can be viewed by two observers with a significant spatial separation
(say from Earth and a distant Solar-orbiting satellite; \citealt{Refsdal1966}) or 
if the event timescale is long enough that Earth moves appreciably in its orbit, then
the microlensing parallax vector $\bm{\pi}_{\rm E}$ \citep{Gould2004, CalchiNovati2016} 
may be measured. Then $\pi_{\rm rel} = \pi_{\rm E} \theta_{\rm E}$, and hence $M$ can be determined.

\section{Observations}

The event OGLE-2016-BLG-1266 (17:51:24.86, -29:44:32.1) J2000.0, galactic coordinates $(l,b) = (-0.04, -1.50)$, was alerted by the Optical Gravitational Lensing Experiment 
\citep[OGLE][]{Udalski2015} on 2016 July 4 UT 11:24, 
based on observations from the 1.3-m Warsaw Telescope at the
Las Campanas Observatory, Chile. The OGLE observations were taken at a cadence of $\sim$ 55 minutes. Photometry of the OGLE
images was extracted using the standard OGLE difference-imaging pipeline.

OGLE-2016-BLG-1266 was also observed by the Korea Microlensing Telescope Network \citep[KMTNet][]{Kim2016} using identical telescopes at 
the Cerro Tololo Inter-American Observatory in Chile; the South African Astronomical Observatory at Sutherland, South Africa;  and the Siding Spring Observatory, Australia.
It was identified as
SAO42T0504.003968 \citep{Kim2017, Kim2018}.
Both the OGLE and KMTNet observations were taken as part of regular surveys, with cadence uninformed by the detection of the event. For KMTNet, the event
is located in two overlapping survey fields, BLG02 and BLG42, giving an effective cadence of $\sim$ 15 minutes. The primary KMTNet observations
were taken in the $I$ band, supplemented by an occasional $V$-band observation.

Photometry was extracted from the KMTNet observations using the software package PYDIA \citep{Albrow2017}, which employs a difference-imaging algorithm based on the modified-delta-basis-function 
approach of \citet{Bramich2013}. The data from field BLG02 observed from SAAO were discarded as they were affected by a cosmetic feature of the detector.
The remaining KMTNet lightcurves were filtered using various image quality criteria and without reference to the lightcurve.

The event was also observed by the {\it Spitzer} space telescope at a wavelength of 3.6 $\mu$m using the IRAC instrument \citep{Fazio2004}. 
These observations were acquired as part of a multi-year project to measure the distances of microlensing planets in the Galaxy  \citep{CalchiNovati2015a, Yee2015}. 
OGLE-2016-BLG-1266 was announced as a {\it Spitzer} target at 2016 July 10 UT 21:15, based on the possibility that it would rise to high magnification, and uploaded to {\it Spitzer} the next day.
The first observation was at UT 18:18 on 16 July. In total, 6 observations were taken during the following 7 days. The sequence of observations was 
terminated at that point due to {\it Spitzer's} Sun-angle restriction. The event was observed for a further 9 epochs by {\it Spitzer} in 2017 after the magnification had fallen to baseline levels. 
{\it Spitzer} photometry was extracted using the methods described in \citet{CalchiNovati2015b}.

\section{Microlensing model from Earth-based observations}

\begin{figure}
\hfill
\includegraphics[width=0.55\linewidth]{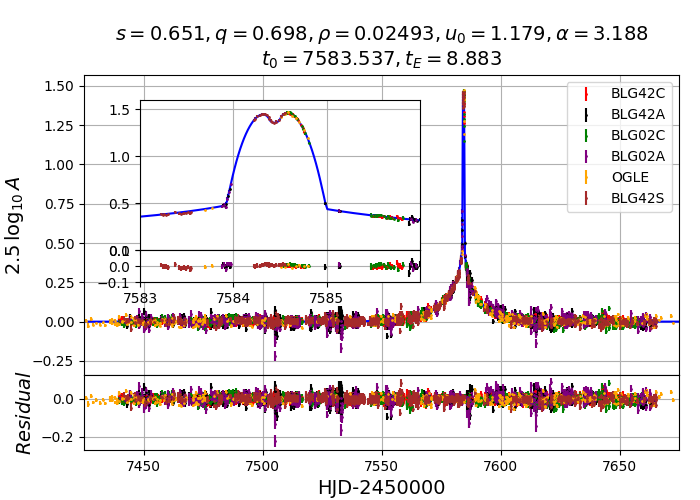}
\hfill
\includegraphics[width=0.4\linewidth]{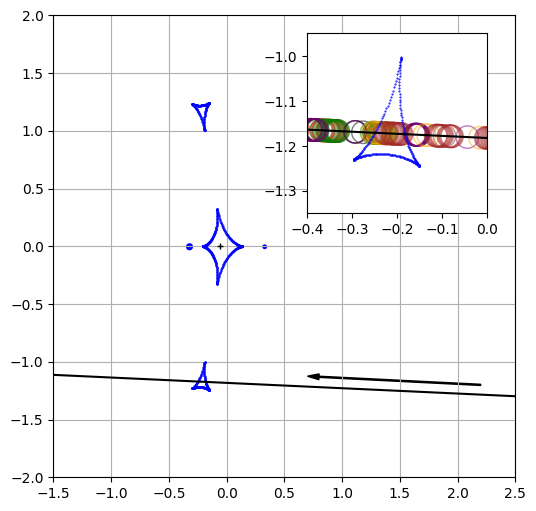}
\caption{Left panel: ground-based lightcurve and model of OGLE-2016-BLG-1266. Right panel:
caustic structure showing the relative source-lens trajectory, which moves from right to left.}
\label{fig:lightcurve}
\end{figure}

The combined ground-based  lightcurve of  OGLE-2016-BLG-1266 is shown in Figure~\ref{fig:lightcurve}. 
It displays a smooth double peak, suggestive of a resolved source crossing a pair of caustics, generated by
a binary lens.

Our analysis of the lightcurve was undertaken using a modified version of the GPU-accelerated code of \citet{McDougall2016}. Initially we performed a search over a fixed
grid of $s, q, r, \alpha$, where $r$ is the distance from the centers of caustics (a reparameterisation of $u_{\rm 0}$).  This established a number of possible  approximate solutions
that were used as starting points for Markov Chain Monte Carlo $\chi^2$ minimization using the EMCEE ensemble sampler \citep{ForemanMackey2013}.
The magnification calculations used the image-centered inverse ray shooting method \citep{Bennett1996, Bennett2010} for locations within three source radii of
a caustic, the hexadecapole approximation \citep{Pejcha2009,Gould2008} for distances between 3 and 30 source radii, and the point-source binary-lens approximation otherwise. 
For the ray shooting calculations, we used a fixed source limb-darkening 
coefficient, $\Gamma = 0.50$, appropriate for the source star color that we derive in the following section. 
From these Markov Chains, a single viable solution was identified at $(s,q) = (0.65, 0.70)$, corresponding to the source passing over one of
the two triangular caustics produced by a close $(s<1)$ binary. 
The corresponding lightcurve and caustic geometry are shown in Figure~\ref{fig:lightcurve}. 
In this paper we plot lightcurves on a scale of $2.5 \log_{\rm 10} A$, where 
$A(t) = (f(t) - f_{\rm B})/f_{\rm S}$, and $f_{\rm B}$ and $f_{\rm S}$ are model-dependent.  
We note that the model implies a small negative blending
for the OGLE data ($f_{\rm B}/f_{\rm S} = -0.044$), equivalent to the flux of an $I_{\rm OGLE} = 20.3$ star. 
As discussed by \citet{Park2004}, 
such low-level negative blending
is a normal feature of microlensing photometry in very crowded bulge
fields.

\section{Source star radius}
\label{Section:source}
 
\begin{figure}
\hfill
\includegraphics[width=0.49\linewidth]{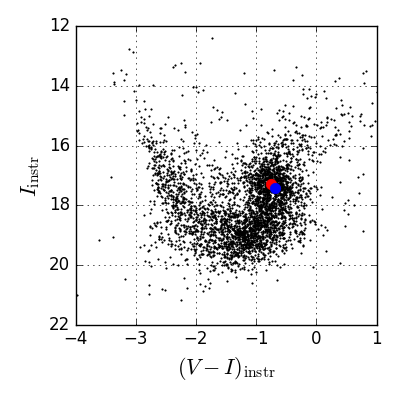}
\hfill
\includegraphics[width=0.49\linewidth]{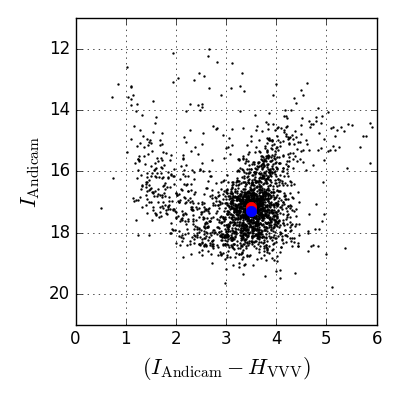}
\caption{Instrumental color-magnitude diagrams of the field of OGLE-2016-BLG-1266.
Left panel: KMT-BLG42C. Right panel: $I_{\rm Andicam}$ and $H_{\rm VVV}$. 
The red clump center
is indicated with a red dot and the deblended source with a blue dot.}
\label{fig:CMD}
\end{figure}

Using  KMTNet CTIO BLG42 images in the $I$ and $V$ bands, we have constructed a DoPHOT 
\citep{Schechter1993}
instrumental color-magnitude diagram (CMD) 
for stars in a 3 arcmin $\times$ 3 arcmin box centered on the event (left panel of Figure~\ref{fig:CMD}). 
From this diagram we measure the red clump centroid to be at $(V-I,I)_{\rm instr} = (-0.76,17.26)$.
From regression of $V$-band flux against $I$-band flux during the event, checked by a two-parameter fit of the $I$-band-determined magnification profile to the $V$-band data,
we determine a deblended instrumental source color $(V-I)_{\rm S,instr} = -0.69 \pm 0.05$ and thus an offset from the clump $\Delta(V-I) = 0.07$.

We have also constructed an instrumental CMD from $I$-band images acquired with the ANDICAM instrument at the 1.3-m CTIO telescope,
and $H$-band catalog measurements of the field from the VVV survey \citep{Saito2012} (right panel of Figure~\ref{fig:CMD}). 
Although $H$-band images were acquired at CTIO simultaneously with the $I$-band images, we opt to 
use VVV measurements for the CMD as they are deeper.
We measure the red clump in this CMD at $(I-H,I)_{\rm instr} = (3.49, 17.16)$. From regression of ANDICAM $I$ and $H$  measurements of the microlensing event,
we determine $(I-H)_{\rm S,Andicam} = -0.63$, which when adjusted for an offset $H_{\rm Andicam} - H_{\rm VVV} = 4.12$ (determined by regression of field stars)
implies an instrumental source color $(I_{\rm Andicam}-H_{\rm VVV})_S = 3.49$ and an offset from the red clump $\Delta(I-H) = 0.00$. 

In principle, color offsets from the Red Clump are filter dependent. However, since our measurement of $\Delta(I-H)$ is essentially zero,
it implies an offset from the clump of zero in any filters. Thus we count this measurement as implying $\Delta(V-I) = 0.00$, and average it
with our previous measurement, $\Delta(V-I) = 0.07$, to obtain a final offset of $\Delta(V-I) = 0.035 \pm 0.05$.

OGLE-2016-BLG-1266 has a galactic longitude close to $0  \deg$,
so from \citet{Nataf2013} and \citet{Bensby2013} we adopt an intrinsic clump centroid  $(V-I,I)_{\rm 0} = (1.06,14.44)$.
From the offset of the deblended source color from the red clump center on the instrumental CMDs, we calculate the intrinsic 
source color $(V-I)_{\rm S,0} = 1.095 \pm 0.05$. Additionally, from the $f_S$ parameter determined from the lightcurve model, we find
$I_S = 17.414$ on the KMT BLG42 instrumental system, and $I_{\rm S,0} =  14.59 \pm 0.05$.

To determine the angular source size, we convert $(V-I)_{\rm S,0}$ to $(V-K)_{\rm S,0} = 2.50 \pm 0.12$ using the empirical color-color equations of \citet{Bessell1988}.
Then, from \citet{Kervella2004}, we find a source angular radius $\theta_* = (5.9 \pm 0.3) \, \mu$as. 

For the same field of stars, using the methods described in \citet{Shvartzvald2017} and \citet{CalchiNovati2015b}, we 
determine that $(I-L)_{\rm S,0} = -3.70 \pm 0.05$ on a {\it Spitzer} system with a 25th magnitude zero point.

From the source angular radius and the lightcurve model we can compute $\theta_{\rm E} = \theta_{*} / \rho = (0.227 \pm 0.011) \, {\rm mas} $. The geocentric lens-source relative proper motion 
is then $\mu_{\rm geo} = \theta_{\rm E} / t_{\rm E} = (9.4 \pm 0.5) \, {\rm mas \,yr^{-1}}$. 

Comparing $\theta_{\rm E}$ and $\mu_{\rm geo}$ with samples from the \citet{Han2003a} model of the Galactic Bulge and Disk, 
Figure 7 in \citet{Penny2016}, (and at this stage ignoring any difference between $\mu_{\rm geo}$ and $\mu_{\rm hel}$) 
we note that  $\mu_{\rm geo}$ is at the extreme of what is possible for a Bulge lens, so the lens
is likely in the Galactic Disk. (See also Figure 1 in \citet{Han2003b}). Our measurement of $\theta_{\rm E}$ implies a total lens mass of 10 $M_{\rm J}$ if the lens
distance is 1.3 kpc and 100 $M_{\rm J}$ if the distance is 5.1 kpc.
  
\section{Parallax Constraints}

The orbit of Earth around the Sun introduces a parallax effect on ground-based observations of microlensing events \citep{Gould1992, Gould2000}. Although
present for all such observations, it is usually only detectable for events with a timescale $t_{\rm E} \gtrsim 30$ d. The effect manifests as a
sinusoidal perturbation on an  otherwise-linear projected source trajectory in the lens plane (see for example \citet{Han2017a, Han2017b, Park2015, Furusawa2013}).

In addition to this annual parallax effect, we fit for
the satellite parallax effect. The {\it Spitzer} telescope is in an Earth-trailing solar orbit, $\sim 95^{o}$  behind Earth in 2016.  
At the time of peak magnification, {\it Spitzer} was located at coordinates (RA,DEC) =  (10:25,09:08) and a distance of 1.484 AU from Earth.
Perpendicular to the direction of OGLE-2016-BLG-1266, the projected distance of {\it Spitzer} from Earth was 
$D_\perp = 1.36$ AU.
When viewed from {\it Spitzer}, the source trajectory across the lens plane is offset by a vector $(\Delta \beta, \Delta \tau)$, in
directions (perpendicular, parallel) to the trajectory observed from Earth. The parallel offset is simply, 
\begin{equation}
\Delta \tau = \frac{t_{\rm 0,Spitzer} - t_{\rm 0,Earth}}{t_{\rm E}},
\end{equation}
but the perpendicular offset suffers from a four-fold satellite parallax degeneracy due to the symmetry of the magnification
field about the lens axis, 
\begin{equation}
\Delta \beta = \pm u_{\rm 0, Spitzer} - \pm u_{\rm 0,Earth},
\label{eqn:satellite_degeneracy}
\end{equation}
as illustrated in \citet{Gould1994}. The sign convention we adopt here is that a positive value of $u_{\rm 0}$ indicates that, during its projected trajectory, the source
approaches the lens on its right hand side.

We make an initial fit to the {\it Spitzer} lightcurve by adopting the ground-based model parameters and exploring a grid in $(\Delta \beta,  \Delta \tau)$
to offset $(t_{\rm 0}, u_{\rm 0})_{\rm Spitzer}$ from $(t_{\rm 0}, u_{\rm 0})_{\rm Earth}$. These constant offset values are used as the reference indices for the $\chi^2$ grid, but
the calculations of the actual model {\it Spitzer} lightcurve use the true offset of each data point at its epoch of observation.

At each point in the grid we map the magnification to the observed {\it Spitzer} flux using Equation~(\ref{eqn:flux}).

To include the {\it Spitzer} source flux constraint derived in the previous section,
we penalise $\chi^2$ with an additional term,
\begin{equation}
\chi^2_{\rm constraint} = \frac{ (2.5 * \log_{\rm 10} (R_{\rm model}/ R_{\rm constraint}))^2 }{ \sigma_{\rm constraint}^2},
\label{eqn:R}
\end{equation}
where $R$ is the $I$-band to $L$-band flux ratio and $\sigma_{\rm constraint}$ is the 
uncertainty in $(I-L)_{\rm S,0}$ \citep{Shin2017}. Mathematically, this is entirely equivalent to a prior on the probability of the model parameters that generate $R_{\rm model}$. We elected not to use the \citet{Shin2017} method of adding an additional penalty for deviations greater than 2-$\sigma$.

The $\chi^2$ grids in $(\Delta \beta,  \Delta \tau)$ for the unconstrained and source-flux-constrained cases are shown in Figure~\ref{fig:delta_beta_tau}.
Comparing the constrained with the unconstrained solutions, it is clear that there is a broad region in $(\Delta \beta,  \Delta \tau)$ space
that is consistent with both the source-flux-constrained and  unconstrained models for the {\it Spitzer} data. 

In the unconstrained case, the lowest-$\chi^2$ solution corresponds to a small region 
at $(\Delta \beta,  \Delta \tau) = (-0.80,-0.34)$, which is not visible in the corresponding flux-constrained map. 
It is instructive to consider the penalty that the source flux
constraint imposes for this particular $(\Delta \beta,  \Delta \tau)$. The $I$-band to $L$-band source-flux ratio
constraint is $R = 1033 \pm 67$ from our CMD analysis. The unconstrained model at that point has $R = 8188$.
It is very unlikely that our $L$-band flux measurement is in error by a factor of 8 (i.e., more than 2 magnitudes).
Thus we consider that this ``best'' of the unconstrained models is ruled out by the $L$-band measurement, and we
only consider the constrained models from here on.

To explore the identified solution space,  
we have run full EMCEE Markov Chains incorporating the standard binary microlensing model with two parallax parameters
$(\pi_{\rm E,E}, \pi_{\rm E,N})$ for the combined ground-based and {\it Spitzer} data, incorporating the {\it Spitzer} source flux constraint. Chains seeded from various points in $(\Delta \beta,  \Delta \tau)$ space all converged to 
one of the two points indicated with plus signs in Figure~\ref{fig:delta_beta_tau}. 

Solution A (``green plus'') corresponds to trajectories for which the six 2016 {\it Spitzer} data points are
located to the left of the central caustic in Figure~\ref{fig:lightcurve}, interior to (i.e., closer to the lens axis than) the
Earth-viewed trajectory. In this region, the magnification is declining smoothly with a steeper slope than the ground-based model.
The $\chi^2$ minimum is located in this solution region at $(\Delta \beta,  \Delta \tau) = (-1.2, 0.6)$. 

Solution B (``yellow plus") is located slightly exterior to the Earth-viewed trajectory.
The lightcurve corresponding to this {\it Spitzer}-viewed trajectory incorporates a small peak 
close to the first data point due to a high magnification region in an extension of a cusp from the planetary caustic.
The $\chi^2$ minimum for this solution is located at $(\Delta \beta,  \Delta \tau) = (0.2, -0.06)$, and is
disfavored relative to Solution A by $\Delta \chi^2 = 17$.

The microlensing parallax, ${\bm \pi}_{\rm E}$, depends on 
$(\Delta \beta,  \Delta \tau)$ as 
\begin{equation}
{\bm \pi}_{\rm E} = \frac{{\rm AU}}{D_\perp} (\Delta \beta, \Delta \tau),
\end{equation}
where $D_\perp$ is the distance of {\it Spitzer} from Earth, perpendicular 
to the line of sight to the lens. At the peak of the event, $D_\perp = 1.36 \, {\rm AU}$, so that
Solution A (``green plus") has  $\pi_{\rm E} \approx 0.98$, while 
Solution B (``yellow plus") has $\pi_{\rm E} \approx 0.17$.

\section{Satellite degeneracy}

\begin{figure}
\hfill
\subfigure{\includegraphics[width=0.49\linewidth]{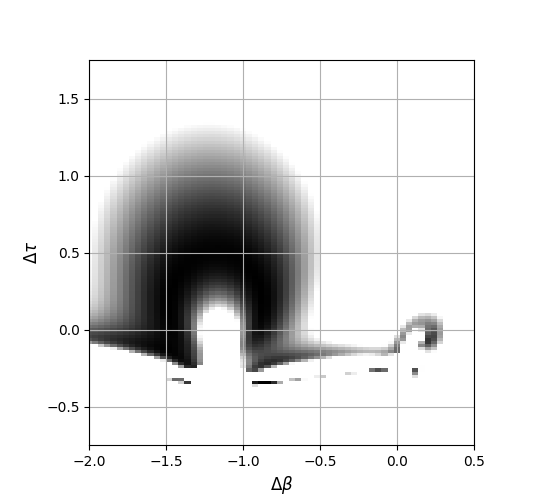}}
\hfill
\subfigure{\includegraphics[width=0.49\linewidth]{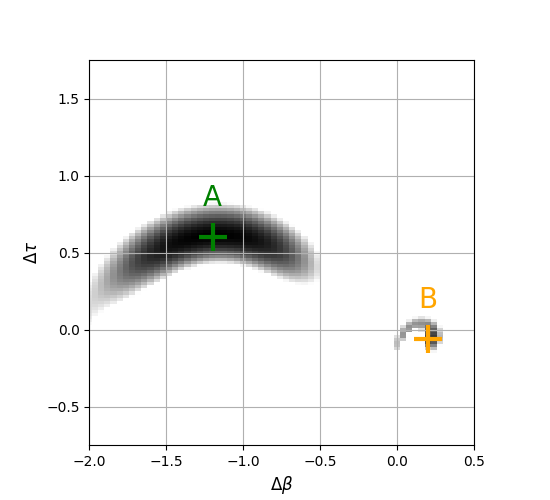}}
\hfill
\caption{Grayscale maps of $\chi^2$ for the unconstrained (left) and flux-constrained (right) fits of the ground-based model offset in  $(\Delta \beta, \Delta \tau)$ to the {\it Spitzer} flux measurements. 
The white level (high cut) is set at $\Delta \chi^2 = 100$ above the minimum $\chi^2$ (black) in each case.
For the flux-constrained case, full MCMC models incorporating the ground-based and {\it Spitzer} data converged to the two
solutions indicated with plus signs.}
\label{fig:delta_beta_tau}
\end{figure}

As discussed at the beginning of the previous section, there exists a four-fold degeneracy in
$(u_{\rm 0, Earth},u_{\rm 0, Spitzer})$ (Equation (~\ref{eqn:satellite_degeneracy})). 
To further investigate the satellite degeneracy, 
we adopted the A (green) and B (yellow) source-flux-constrained solutions from Figure~\ref{fig:delta_beta_tau} 
and ran EMCEE chains to explore the  complete set of  $(\pm u_{\rm 0, Earth},\pm u_{\rm 0, Spitzer})$ solution regions.
 
For Solution A at $(\Delta \beta,  \Delta \tau) = (-1.16, 0.60)$
(green plus symbol in the right panel of Figure~\ref{fig:delta_beta_tau}), the $+u_{\rm 0, Spitzer}$ and $-u_{\rm 0, Spitzer}$
trajectories lie along the lens axis and are almost identical for $+u_{\rm 0, Earth}$. We refer to these as the A  (``green'') solutions.

In contrast, for Solution B (yellow plus symbol in the right panel of figure Figure~\ref{fig:delta_beta_tau}) at $(\Delta \beta,  \Delta \tau) = (-0.2,0.06)$, there is a separate $-u_{\rm 0, Spitzer}$
trajectory that lies above the upper triangular caustic (and was outside the range of the $(\Delta \beta,  \Delta \tau)$
grid search). 
We refer to these solutions as the B (``yellow") solutions.

\begin{figure}
\hfill
\subfigure{\includegraphics[width=0.48\linewidth]{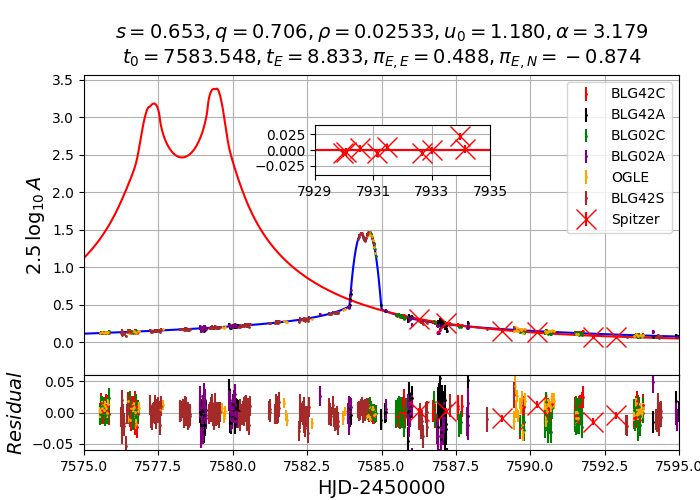}}
\hfill
\subfigure{\includegraphics[width=0.48\linewidth]{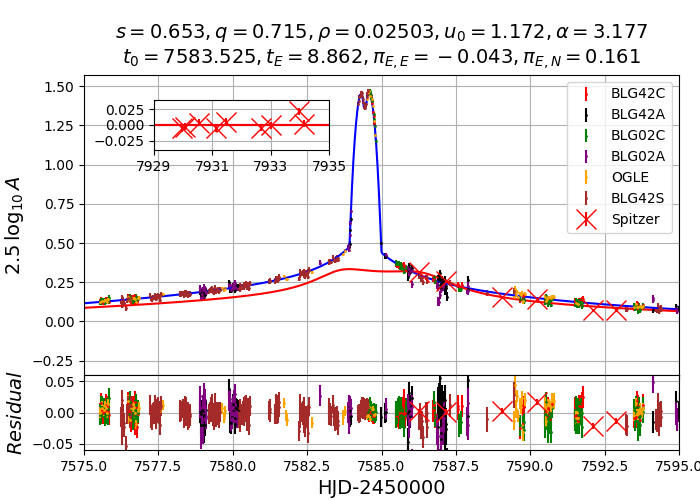}}
\hfill
\caption{Lightcurves for the Solution A  ``green plus" (left) and Solution B ``yellow plus" (right) source-flux-constrained models from Figure~\ref{fig:delta_beta_tau}. Inset plots show the 2017 {\it Spitzer} data.}
\label{fig:constrained_solutions_lightcurves}
\end{figure}

Microlensing parameters derived from the eight models are shown in 
the left columns for each geometry in Tables~\ref{table:microlensing_results_green} 
and \ref{table:microlensing_results_yellow}.
Given the apparent degeneracies, the eight solutions correspond to three different 
microlensing parallaxes; small-parallax B ($\pi_{\rm E} \sim 0.17$), large-parallax B ($\pi_{\rm E} \sim 1.8$), and A ($\pi_{\rm E} \sim 0.97$).
Representative lightcurves for the A (green) $+/+$ and B (yellow) $+/+$ geometries are shown in 
Figure~\ref{fig:constrained_solutions_lightcurves}.

 For the solution-A trajectories, there is a 
 small $\chi^2$ difference in the $\pm u_{\rm 0,Earth}$ solutions due to the small annual-parallax-induced 
 curvature in each trajectory, with the $+u_{\rm 0,Earth}$ solution being marginally favored. The 
 $\pm u_{\rm 0,Spitzer}$ solutions for each $u_{\rm 0,Earth}$ are fully degenerate.
 
In contrast, the second set of models (solution B) separate in $\chi^2$ for  the two cases that $u_{\rm 0,Spitzer}$
has the same or opposite sign as $u_{\rm 0,Earth}$ (with the opposite sign models favored by $\Delta \chi^2 \approx 7 - 10$),
but are otherwise degenerate in $\pm u_{\rm 0,Earth}$. 
   
 \section{Lens orbital motion}
 
Ignoring projection effects, a Keplerian orbit for the masses and separation derived in the previous section would have
a period of about 1.6 years. This suggests that lens orbital motion may be a detectable 
 and significant effect. We have thus extended our models  with first order lens motion parameters $\frac{d \alpha}{d t}$ and 
 $\frac{d s}{d t}$. 
 
 We require that the complete set of model parameters are consistent with a bound
 orbit, in particular that the projected kinetic energy be less than the potential energy. 
 From \citet{Dong2009}
 \begin{equation}
 \left( \frac{KE}{PE} \right)_\perp = \frac{ 2 ({\rm AU})^2 }{c^2} \frac{\pi_{\rm E}}{\theta_{\rm E}} 
 \frac{ \left[ \left( \frac{1}{s} \frac{ds}{dt} \right)^2 + \left( \frac{d\alpha}{dt} \right)^2 \right] s^3 }
 { \left[ \pi_{\rm E} + \left( \frac{\pi_s}{\theta_{\rm E}} \right) \right]^3}.
 \end{equation}
 In convenient units, this leads to a constraint,
 \begin{equation}
 \left( \frac{1}{s} \frac{ds}{dt} \right)^2 + \left( \frac{d\alpha}{dt} \right)^2 < 
 (9.644 \, {\rm yr}^{-2}) 
  \left( \frac{\theta_{\rm E}}{{\rm mas}} \right) \frac{1}{ s^3 \pi_{\rm E}} 
 \left( \pi_{\rm E} + \frac{ 1 }{ \left( \frac{D_S}{{\rm kpc}} \right) \left( \frac{\theta_{\rm E}}{{\rm mas}} \right) } \right)^3.
 \end{equation}
 
 From Section~\ref{Section:source}, we assume that the source is at the red-clump distance, 8.18 kpc, 
and that $\theta_{\rm E} = 0.227 \pm 0.011 \, {\rm mas}$. We implement the constraint as a prior, with the hard upper boundary
softened by the uncertainty in $\theta_{\rm E}$.

\begin{figure}
\hfill
\subfigure{\includegraphics[width=0.48\linewidth]{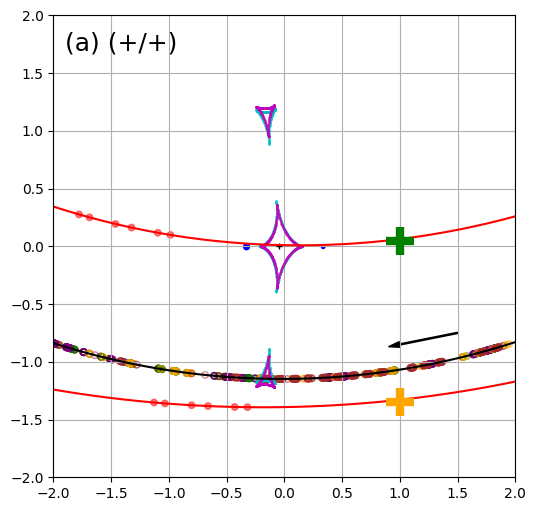}}
\hfill
\subfigure{\includegraphics[width=0.48\linewidth]{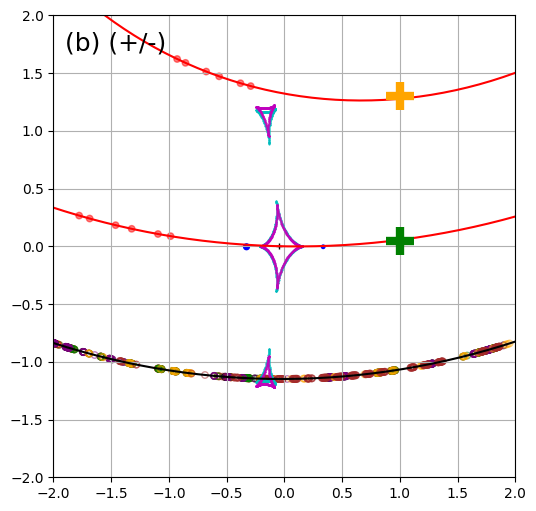}}
\hfill
\subfigure{\includegraphics[width=0.48\linewidth]{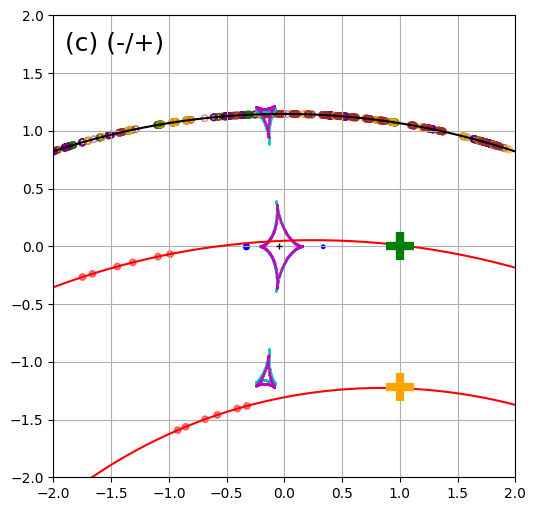}}
\hfill
\subfigure{\includegraphics[width=0.48\linewidth]{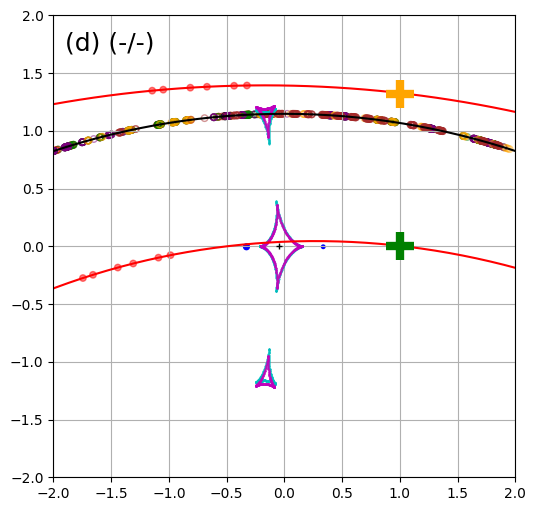}}
\hfill
\caption{Effective source trajectories relative to the lens for the final ground-based and {\it Spitzer} models incorporating
parallax and lens orbital motion. The four panels show the different satellite-degenerate geometries. 
The Earth-viewed source trajectory is in black, and the {\it Spitzer}-viewed trajectories
for the Solution A (green) and Solution B (yellow) series of solutions are in red. 
All source trajectories are from right to left, and the circles 
indicate data epochs. The caustics are shown in cyan/magenta at $\Delta t = t_{\rm E}$ before/after  HJD 2457584.44, the 
epoch when the ground trajectory is at the center of the caustic.}
\label{fig:caustics}
\end{figure}

We have run models seeded from the eight satellite-parallax degenerate solution regions discussed above.
The resulting parameters are listed in the right hand columns for each geometry in Tables~\ref{table:microlensing_results_green} and \ref{table:microlensing_results_yellow},
and the geometries are displayed in Figure~\ref{fig:caustics}.

Chains for solution A (green) converge to almost the same solution, again with a small $\chi^2$ difference
of $\sim$ 4 between the $\pm u_{\rm 0,Earth}$ solutions. The overall $\chi^2$ is lowered by $\sim$ 12 relative to
the models without lens orbital motion. 

Again, as expected, the solution-B models (yellow) converge to 
different solutions for $\pm u_{\rm 0,Spitzer}$. 
The $+/-$ and $-/+$ solutions have $\chi^2 \approx 11 - 16$ smaller than the  $+/+$ and $-/-$ solutions. Relative to the 
$+/-$ solution-A model, the best of these is disfavored by
$\Delta \chi^2 \approx 7$.

\begin{figure}
\includegraphics[width=\columnwidth]{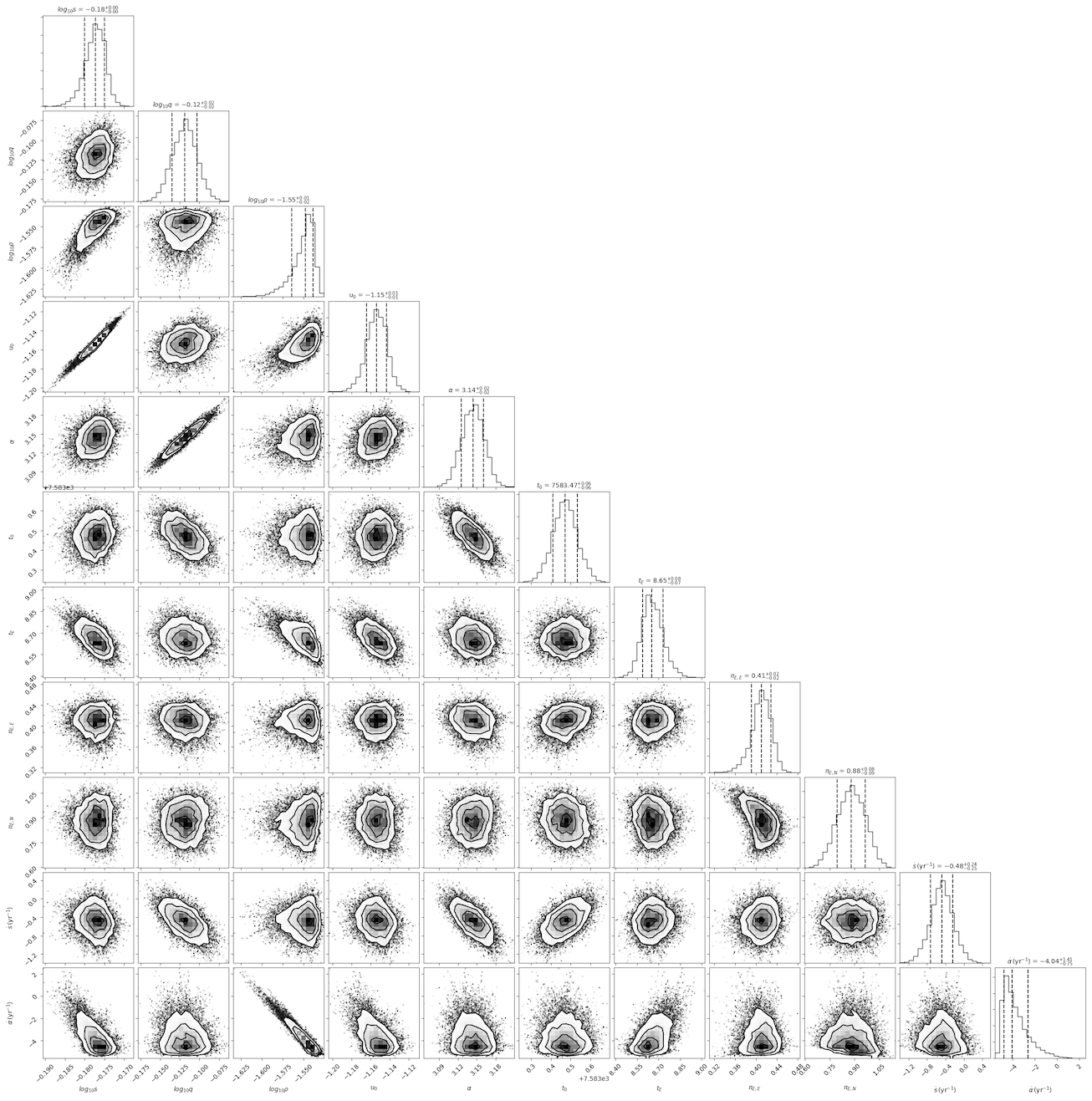}
\caption{Two-dimensional covariance plots for the MCMC samples for the 11 parameters in the A (green) $-/-$ solution. Contours are drawn at (0.5, 1.0, 1.5, 2.0)-sigma. The cutoff apparent in the lower row of panels for parameter $\dot{\alpha}$ and the third column for parameter $\log_{\rm 10} \rho$ is 
due to the orbital kinetic energy constraint.}
\label{fig:triangle_plots}
\end{figure}

In Figure~\ref{fig:triangle_plots} we show the posterior parameter distribution for the solution A $-/-$
model (which ultimately becomes out favored model in Section 8 below). 
Corresponding plots for the other 7  models are similar. 
The effect of the kinetic energy prior is apparent in $\frac{d \alpha}{d t}$ and $\rho$.
If it were not for this physical constraint, the data would force $\frac{d \alpha}{d t}$ to $7 \pm 2$.
The kinetic energy prior has an effect in all cases, but each 
energy-constrained solution is always part of the same $\chi^2$ minimum 
as a corresponding unrestricted (non-physical) model. 

Overall, the inclusion of lens orbital motion in the models changes
slightly the other parameters, and improves $\chi^2$ slightly for all models.

 Both of the $\pm u_{0,Earth}$  A solutions imply that the lens is a binary 
 with component masses of  $\sim$ 16 and 12 Jupiter masses. 
 The higher mass component is a brown dwarf, and the
 lower mass component is on the dividing line between a brown dwarf and a super-Jupiter planet (sub-brown dwarf).
 The lens is located at a distance of 3.0 kpc from Earth, and the 
 components have a projected separation of 0.4 AU. 
 
 The ``same-sign'' B solutions (with $\Delta \chi^2 = 17$ relative to the A solutions) are for a 90 and 68 Jupiter-mass binary at 6.2 kpc with
 a projected separation of 0.9 AU, while the ``opposite-sign'' B solutions (with $\Delta \chi^2 = 8$ relative to the A solutions) 
 are for an 8 and 6 Jupiter-mass binary at 2.1 kpc with
 a projected separation of 0.3 AU.

\section{Which solution is correct?}

In this section we use several lines of evidence to assess the solutions obtained above. Our approach is 
similar to that of \citet{CalchiNovati2017}.

\subsection{The best lightcurve fit}

In Tables~\ref{table:microlensing_results_green} and~\ref{table:microlensing_results_yellow} 
we list $\chi^2$ for each fit. In all cases, we have found that the six {\it Spitzer} data points from 2016 comprise the major source
of $\Delta \chi^2$. Irrespective of whether lens orbital motion is included in the models, the A (green) series
of solutions have the best formal fit, with the $+u_{\rm 0,Earth}$ models being slightly better than the $-u_{\rm 0,Earth}$ models. 
The B (yellow) series of solutions
are less favored, but not by a large amount. Formally, the probability of each solution relative to the best one
is lowered by $e^{-\Delta \chi^2 / 2}$, so that the best of the large-parallax B solutions has a probability of only 
0.040 relative to the A solution (i.e., a 2.5-$\sigma$ difference), and the best small-parallax B solution 
has a relative probability of $5.33 \times 10^{-5}$ (4.4-$\sigma$).
However, these formal probabilities rest on the assumption that all data are independent
and Gaussian-distributed, and that data uncertainties are accurate. Such conditions are never satisfied for 
microlensing photometry. On this basis, we are unable to
reject entirely  the yellow solutions.

\subsection{The Rich argument}

 The ``Rich argument" is elucidated in \citet{CalchiNovati2015a}. Briefly, for a point-lens microlensing event, 
 when considering two alternate satellite-degenerate
 solutions from the same $u_{\rm 0,Earth}$ model, the one with the smallest parallax is usually correct. This is because if
 the true parallax solution is small, it will always generate a large-parallax counterpart. However, if the true parallax is large, then there is a much smaller probability, $(\pi_{\rm E,small}/\pi_{\rm E,large})^2$,  that the parallax of the counterpart 
solution is small. This probability factor is based on the rotational symmetry of the magnification field about
the lens. For incomplete satellite lightcurves, the true probability factor can be larger because a two-parameter
fit can map different magnification patterns to the same flux lightcurve. 
 
For binary lenses, the geometric degeneracy on which the Rich argument is based, exists only for 
cases in which the source trajectory is almost parallel to the lens axis, as is the situation we
are considering here. The two B (yellow) solutions represent an analogous situation
to the large- and small-parallax solutions for a point lens. However, we cannot naively
apply the point-lens relative probability factor because the magnification pattern for a binary lens 
does not have the rotational symmetry of the single-lens pattern, and our {\it Spitzer} lightcurve does
not have full coverage.

The A-solution degeneracy with either of the B solutions is not
a true geometric degeneracy. It exists because of our limited epochs of {\it Spitzer}
observations and would not be present if we had full temporal coverage.

We have assessed the relative probabilities of the various solutions by simulating
{\it Spitzer} lightcurves for the three different parallax amplitudes ($\pi_{\rm E}$) and for different 
angles ($\omega$) with respect to the source trajectory. For each simulation, we 
held the ground-determined microlensing parameters constant, and
computed a flux lightcurve by combining the previously-determined {\it Spitzer} source and blend flux with the
the magnification, $A(t | \pi_{\rm E}, \omega)$ at the {\it Spitzer} epochs. We then added the residuals of the {\it Spitzer}
data relative to the A $(+/+)$ model fit. 

To determine the probability factor for the large-parallax B $(+/-)$ solution relative to the small-parallax
B $(+/+)$ solution, we have simulated lightcurves for 360 values of $\omega$ for the B $(+/+)$ parallax
amplitude. For each of these we have made a two-parameter, source-flux constrained fit using the magnification at the constant
B $(+/-)$ parallax amplitude, found from the transformation 
$\Delta \beta_{\rm alt} = - \Delta \beta - 2 u_{\rm 0, Earth}$, and allowing the angle $\omega_{\rm alt}$
to vary. For each $\omega$, we accumulate the $\Delta \chi^2$ between the large-parallax and small-parallax
fits. We then compute the probability that a true small parallax would have a large parallax degeneracy as being
the fraction of angles $\omega$ for which  $\Delta \chi^2$ is less than some threshold value. We then repeat this exercise
in reverse, generating a set of large-parallax simulations, and finding the small-parallax fits.  
The ratio of these two probabilities then gives the {\it a-priori} probability of a large parallax solution relative to 
a small one for the given geometry, satellite observation epochs, source-flux constraints, and
observation residuals, 
independent of the actual measured satellite flux values. 
To simplify the interpretation, we have made these synthetic lightcurves and fits without
including the effects of lens orbital motion.

We adopt a threshold $\Delta \chi^2 = 20$ 
similar to the range of actual measured $\Delta \chi^2$ for our different degenerate solutions discussed in 
the previous sections. We find that there is a 0.43 probability that a true small B parallax would generate a
large B parallax degenerate solution. If the source trajectory were exactly parallel to the lens axis, we would
expect this factor to be exactly 1.0. (In the current geometry, the factor rises to unity were we to increase our
$\Delta \chi^2$ threshold to 80.) We find that there is a 0.40  probability that a true large B parallax can generate a
small B parallax degenerate solution. This probability is much larger than what would be the case
for continuous {\it Spitzer} lightcurve coverage. Combined, we find an overall probability 
of 0.935 of a large B parallax relative to a small B parallax.

To determine the probability factor for the A $(+/+)$ model relative to the small-parallax B $(+/+)$ 
model we carry out a similar calculation, but use the A $(+/+)$ model parallax amplitude rather than
the parallax of the alt solution. From these calculations we find that there is a 0.33 probability that
a true small B parallax would generate a degenerate solution with the parallax amplitude of the A model,
a 0.37 probability that a true solution with the parallax amplitude of the A model would generate a
degenerate small-parallax B solution, and an overall probability of 1.11 of an A-model parallax relative
to a  small-parallax B model.

In summary, we have found that the overall {\it a-priori} relative probability factors stemming from this specific geometry
and set of {\it Spitzer} observation epochs are close to unity, so have little effect on our relative
assessment of the different solutions.

\subsection{Galactic rotation}

For a flat rotation curve, and from the Local Standard of Rest (LSR) perspective, the rotational component 
of the proper motion of a disk star interior to the Sun's orbit and relative to the Bulge, 
$\mu_{\rm rot, rel} = v_{\rm rot,disk}/D_{\rm bulge}$, is independent of the star's distance. 
Projected onto the lens plane, the disk of the Milky Way rotates in a direction $59.3 \deg$ North of East. 
In the absence of random velocity dispersions for the disk and Bulge, 
we would expect disk lenses to have a relative proper motion
in this direction if we were observing from the LSR. 
Adopting $v_{\rm rot,disk} = 235 \, {\rm km \, s^{-1}}$, and a 
Bulge distance of 8.18 kpc, gives $\mu_{\rm rot, rel} =  6.06 \, {\rm mas \, yr^{-1}}$.
Added to this overall disk rotation, individual disk stars have a velocity dispersion, $\sigma_{\rm disk}$. Given its low
galactic latitude, the lens in OGLE-2016-BLG-1266 is almost certainly part of the old thin disk,
for which  $\sigma_{\rm disk} \approx 15 \, {\rm km \, s}^{-1}$.

\begin{figure}
\includegraphics[width=0.5\linewidth]{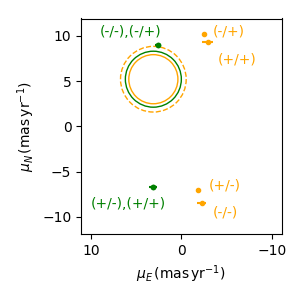}
\caption{Proper motion of the different solutions in the local standard of rest frame of reference with the
same color-coding as used in previous figures.
Circles show the 1-$\sigma$ expected distribution of relative lens-source proper motions for a disk lens
at the distance of each solution. The yellow solid circle applies to the B-series $(+/+)$ and $(-/-)$ solutions. The yellow 
dashed circle applies to the B-series $(+/-)$ and $(-/+)$ solutions.}
\label{fig:proper_motion}
\end{figure}

From its location on the CMD, we assume our source star is part of the Bulge population.
We have adopted $\sigma_{\rm bulge} = 100  \, {\rm km \, s^{-1}}$, as an average of the Y and Z direction
Bulge velocity dispersions from \citet{Bland-Hawthorn2016}.
To compare the relative lens-source proper motion for our various models with that expected for 
disk lenses, we transform to the LSR by adding the projection of Sun's peculiar
velocity, $(V,W) = (12.24, 7.25)  \, {\rm km \, s}^{-1}$, to the 
relative lens-source proper motion. The resultant
LSR relative lens-source proper motions, $\bm{\mu}_{\rm LSR}$ for each model, are shown in Figure~\ref{fig:proper_motion}
along with the galactic expectation, $\bm{\mu}_{\rm MW}$ with a dispersion 
$\sigma_{\rm \mu, MW}^2 =  \sigma^2_{\rm disk}/D_L^2 + \sigma^2_{\rm bulge}/D_S^2$.

We can see that the A (green) solutions for $-u_{\rm 0,Earth}$ are well aligned with Galactic Disk rotation.
The B (yellow) solutions for $+u_{\rm 0,Spitzer}$ are also plausible, but the remaining models are rather 
improbable.

For each model solution, we can form a probability that the lens has the expected proper motion of the 
Galactic Disk,  
\begin{equation}
P_{\rm pm} = \exp^{\biggl( -{ \left| \bm{\mu}_{\rm MW} - \bm{\mu}_{\rm LSR} \right|^2 \over 2 (\sigma_{\rm \mu, MW}^2 + \sigma_{\rm \mu, LSR}^2) } \biggr) },
\end{equation}
(see Tables~\ref{table:microlensing_results_green} and~\ref{table:microlensing_results_yellow}).
Based on their proper motion correspondence with our galactic rotation model, the B (yellow)
$+u_{\rm 0,Spitzer}$ models are  4 - 15 times less probable than the A (green) $-u_{\rm 0,Earth}$ models.

\subsection{Combined probability}

We have discussed three factors that we consider important for assessing the relative
merits of the A and B series of models for OGLE-2016-BLG-1266. For each of these,
we can assign a relative probability; $P_{\rm lightcurve, rel} = e^{-(\chi^2 - \chi^2_{\rm best})/2}$ from
the lightcurve fits, $P_{\rm Rich}$ from the Rich argument, and $P_{\rm pm}$ from the
proper motion correspondence to galactic rotation. We have multiplied the three probabilities for each model
and normalised to the maximum to compute a net relative probability, $P_{\rm total, rel}$, also listed in 
Tables~\ref{table:microlensing_results_green} and~\ref{table:microlensing_results_yellow}.

Considering all factors, the most-favored solutions are the degenerate A-series $-u_{\rm 0,Earth}$
models. These models imply a 16-$M_{\rm J}$ + 12-$M_{\rm J}$
mass lens at a distance of 3.1 kpc. We note that these solutions have a significantly larger parallax
($\pi_{\rm E} = 0.98)$ than any of the previous events measured by {\it Spitzer}.
(The next largest parallax is OGLE-2016-BLG-1195 with $\pi_{\rm E} = 0.45$; \citealt{Shvartzvald2017}.)

Relative to the A solutions, the best of the B solutions is the large-parallax
$(-/+)$ model with a relative probability of $1.47 \times 10^{-2}$.  
The small-parallax $(+/+)$
model has a relative probability of $7.61 \times 10^{-6}$. 
Respectively, these models imply an 8-$M_{\rm J}$ + 6-$M_{\rm J}$ mass lens at a distance of 2.0 kpc
and a 90-$M_{\rm J}$ +  70-$M_{\rm J}$ pair at 6.2 kpc. The relative probabilities correspond to 2.91-$\sigma$
and 4.86-$\sigma$ differences from the favored model.

\subsection{Is the favored lens mass plausible?}

The initial mass function (IMF) for brown dwarfs below $M$ = 0.1 M$_\sun$ ($105 \, M_{\rm J}$) is not well established, and may depend 
on environment. The \citet{Kroupa2002} and \citet{Chabrier2003} IMFs for single objects increases toward lower mass in this range,
but there is evidence that the MF ``turns over" at increasingly higher masses with age in stellar clusters \citep{Chabrier2003}.
There is little evidence of a large
decline between 0.1 M$_\sun$ and 0.01 M$_\sun$ in the studies of  \citet{Alves2013, Jeffries2012, Gagne2017}. 
Overall, there is little reason from the lens-mass results to reject the high-parallax low-mass model in favor of the 
lower-parallax higher-mass one.

\subsection{Falsification of the favored model}

Our adopted model for OGLE-2016-BLG-1266 is for a 16-$M_{\rm J}$ + 12-$M_{\rm J}$
mass lens at a distance of 3.1 kpc. Given its low mass, we do not expect the
lens to be directly observable with any presently conceived instruments. 
This is also the case for the B $(-/+)$ model, which corresponds to a binary composed of two
planetary-mass objects.

However, the more massive of the plausible  
challenger models (B $+/+$)
consists of a 90-$M_{\rm J}$ + 70-$M_{\rm J}$ mass lens at a distance of 6.2 kpc. 
From \citet{Dupuy2017},we expect that this pair would have absolute $J$ and $K$ magnitudes of 11 and 10, and
so apparent magnitudes $J \sim 25$, $K \sim 24$. This solution has a heliocentric lens-source relative proper motion
$(\mu_{\rm N},\mu_{\rm E})$ of $(8.68,-2.94)$  mas yr$^{-1}$. In ten years there would be a separation of
 92 mas between the lens and the $K=13.2$ source in the indicated direction. Resolving the lens and source 
 for the high-mass  model should be 
 within the first-light capability of diffraction-limited near-infrared imagers on the coming generation of extremely large telescopes, for example ELT-CAM on E-ELT.

\section{Discussion}

Objects like OGLE-2016-BLG-1266 challenge our understanding of what is meant by a planet. If the low-mass component of our favored 
model were associated with a star, or a brown dwarf with significantly higher mass, then it would be
be described as a planet. However, with masses so close, both components of the binary may instead belong to the 
very low-mass end of the stellar initial mass function. 
We note that the survey of \citet{Mroz2017} has identified several
short-time-scale binary events that may be part of such a population.

For several reasons, detecting single planetary mass objects  (often referred to as ``free-floating planets") by the microlensing method is a more difficult task than 
detecting binary lenses. Firstly, the peak magnification is generally lower and thus will have
a lower probability of detection as a microlensing event.
Second, the mass of single lenses can only be inferred from a measurement of $t_{\rm E}$, and that parameter is extremely degenerate with blending, and subject to 
incorrect inference if derived from lightcurve data with any systematic correlation between neighboring points.
Third, it is difficult to establish a microlensing parallax measurement for short-$t_{\rm E}$ events, because for Earth-orbital
parallax measurements the trajectory of Earth does not deviate much from linear during the event duration, and
for satellite parallaxes it is difficult to target satellite observations while the event is still significantly magnified. 

There are currently four published single-lens events with secure lens-mass measurements from 
{\it Spitzer}  \citep{Zhu2016, Chung2017, Shin2018}, and two from ground-only measurements
\citep{Gould2009, Yee2009}, all with masses in the brown-dwarf regime.

There are currently no secure detections of single planetary mass objects
by the microlensing method \citep{Mroz2017}.
The object very recently detected by \citet{Mroz2018} may be the first isolated ``planet", but even for that event, the
presence of a stellar host at a separation $\geq 15$ AU cannot be ruled out.

Single planetary-mass lenses may be found
more readily in the future with the advent of the WFIRST mission, which will observe Galactic Bulge microlensing events with
high photometric precision and less blending than from the ground. \citet{Gould2016} shows how the presence of
stellar companions to such single-lens candidates can be detected or ruled out by WFIRST and ground-based 
adaptive-optics observations.
We should always bear in mind that, in the absence of an evolutionary history, the designation 
of low-mass single lenses as free-floating ``planets" may be incorrect.

\section{Summary}

Using data from the KMTNet and OGLE telescopes, and the {\it Spitzer} satellite, we have analysed the
microlensing event OGLE-2016-BLG-1266. Our models show that the lens is very likely composed of a
16-$M_{\rm J}$ + 12-$M_{\rm J}$ binary at a distance of 3.1 kpc.

Two alternative models are unlikely, but cannot be entirely rejected. One of these models corresponds
to a 6-$M_{\rm J}$ + 8-$M_{\rm J}$ ``planet-planet'' binary at a distance of 2.0 kpc. The second of these alternatives, a 70-$M_{\rm J}$ + 90-$M_{\rm J}$ binary at
6.2 kpc, would be directly observable with the next generation of telescopes and instrumentation.

\section{Acknowledgements}

MDA is supported by the Marsden Fund under contract UOC1602, and is grateful for
the award of an ESO Visiting Fellowship in December 2017 / January 2018 during which time
this paper was completed.
Work by WZ, YKJ, and AG were supported by AST-1516842 from the US NSF.
WZ, IGS, and AG were supported by JPL grant 1500811.
Work by C.H. was supported by the grant (2017R1A4A1015178) of
National Research Foundation of Korea.
This research has made use of the KMTNet system operated by the Korea
Astronomy and Space Science Institute (KASI) and the data were obtained at
three host sites of CTIO in Chile, SAAO in South Africa, and SSO in
Australia.
The OGLE project has received funding from the National Science Centre,
Poland, grant MAESTRO 2014/14/A/ST9/00121 to AU.
Work by YS was supported by an appointment to the NASA Postdoctoral
Program at the Jet Propulsion Laboratory, California Institute of
Technology, administered by Universities Space Research Association
through a contract with NASA.
This work is based in part on observations made with the Spitzer Space Telescope, which is operated by the Jet Propulsion
Laboratory, California Institute of Technology under a contract with NASA. 


\begin{thebibliography}{}

\expandafter\ifx\csname natexlab\endcsname\relax\def\natexlab#1{#1}\fi
\providecommand{\url}[1]{\href{#1}{#1}}

\bibitem[{Albrow(2017)}]{Albrow2017}
Albrow, M. 2017, MichaelDAlbrow/pyDIA: Initial release on github., , ,
  doi:10.5281/zenodo.268049.
\newblock \url{https://doi.org/10.5281/zenodo.268049}

\bibitem[{{Alves de Oliveira} {et~al.}(2013){Alves de Oliveira}, {Moraux},
  {Bouvier}, {Duch{\^e}ne}, {Bouy}, {Maschberger}, \& {Hudelot}}]{Alves2013}
{Alves de Oliveira}, C., {Moraux}, E., {Bouvier}, J., {et~al.} 2013, \aap, 549,
  A123

\bibitem[{{Baraffe} {et~al.}(2008){Baraffe}, {Chabrier}, \&
  {Barman}}]{Baraffe2008}
{Baraffe}, I., {Chabrier}, G., \& {Barman}, T. 2008, \aap, 482, 315

\bibitem[{{Bate}(2012)}]{Bate2012}
{Bate}, M.~R. 2012, \mnras, 419, 3115

\bibitem[{{Bennett}(2010)}]{Bennett2010}
{Bennett}, D.~P. 2010, \apj, 716, 1408

\bibitem[{{Bennett} \& {Rhie}(1996)}]{Bennett1996}
{Bennett}, D.~P., \& {Rhie}, S.~H. 1996, \apj, 472, 660

\bibitem[{{Bensby} {et~al.}(2013){Bensby}, {Yee}, {Feltzing}, {Johnson},
  {Gould}, {Cohen}, {Asplund}, {Mel{\'e}ndez}, {Lucatello}, {Han}, {Thompson},
  {Gal-Yam}, {Udalski}, {Bennett}, {Bond}, {Kohei}, {Sumi}, {Suzuki}, {Suzuki},
  {Takino}, {Tristram}, {Yamai}, \& {Yonehara}}]{Bensby2013}
{Bensby}, T., {Yee}, J.~C., {Feltzing}, S., {et~al.} 2013, \aap, 549, A147

\bibitem[{{Bessell} \& {Brett}(1988)}]{Bessell1988}
{Bessell}, M.~S., \& {Brett}, J.~M. 1988, \pasp, 100, 1134

\bibitem[{Best {et~al.}(2017)Best, Liu, Dupuy, \& Magnier}]{Best2017}
Best, W. M.~J., Liu, M.~C., Dupuy, T.~J., \& Magnier, E.~A. 2017, The
  Astrophysical Journal Letters, 843, L4.
\newblock \url{http://stacks.iop.org/2041-8205/843/i=1/a=L4}

\bibitem[{{Bland-Hawthorn} \& {Gerhard}(2016)}]{Bland-Hawthorn2016}
{Bland-Hawthorn}, J., \& {Gerhard}, O. 2016, \araa, 54, 529

\bibitem[{{Boss} {et~al.}(2007){Boss}, {Butler}, {Hubbard}, {Ianna},
  {K{\"u}rster}, {Lissauer}, {Mayor}, {Meech}, {Mignard}, {Penny},
  {Quirrenbach}, {Tarter}, \& {Vidal-Madjar}}]{Boss2007}
{Boss}, A.~P., {Butler}, R.~P., {Hubbard}, W.~B., {et~al.} 2007, Transactions
  of the International Astronomical Union, Series A, 26, 183

\bibitem[{{Bramich} {et~al.}(2013){Bramich}, {Horne}, {Albrow}, {Tsapras},
  {Snodgrass}, {Street}, {Hundertmark}, {Kains}, {Arellano Ferro}, {Figuera},
  \& {Giridhar}}]{Bramich2013}
{Bramich}, D.~M., {Horne}, K., {Albrow}, M.~D., {et~al.} 2013, \mnras, 428,
  2275

\bibitem[{{Calchi Novati} \& {Scarpetta}(2016)}]{CalchiNovati2016}
{Calchi Novati}, S., \& {Scarpetta}, G. 2016, \apj, 824, 109

\bibitem[{{Calchi Novati} {et~al.}(2015{\natexlab{a}}){Calchi Novati}, {Gould},
  {Udalski}, {Menzies}, {Bond}, {Shvartzvald}, {Street}, {Hundertmark},
  {Beichman}, {Yee}, {Carey}, {Poleski}, {Skowron}, {Koz{\l}owski}, {Mr{\'o}z},
  {Pietrukowicz}, {Pietrzy{\'n}ski}, {Szyma{\'n}ski}, {Soszy{\'n}ski},
  {Ulaczyk}, {Wyrzykowski}, {OGLE Collaboration}, {Albrow}, {Beaulieu},
  {Caldwell}, {Cassan}, {Coutures}, {Danielski}, {Dominis Prester},
  {Donatowicz}, {Lon{\v c}ari{\'c}}, {McDougall}, {Morales}, {Ranc}, {Zhu},
  {PLANET Collaboration}, {Abe}, {Barry}, {Bennett}, {Bhattacharya},
  {Fukunaga}, {Inayama}, {Koshimoto}, {Namba}, {Sumi}, {Suzuki}, {Tristram},
  {Wakiyama}, {Yonehara}, {MOA Collaboration}, {Maoz}, {Kaspi}, {Friedmann},
  {Wise Group}, {Bachelet}, {Figuera Jaimes}, {Bramich}, {Tsapras}, {Horne},
  {Snodgrass}, {Wambsganss}, {Steele}, {Kains}, {RoboNet Collaboration},
  {Bozza}, {Dominik}, {J{\o}rgensen}, {Alsubai}, {Ciceri}, {D'Ago},
  {Haugb{\o}lle}, {Hessman}, {Hinse}, {Juncher}, {Korhonen}, {Mancini},
  {Popovas}, {Rabus}, {Rahvar}, {Scarpetta}, {Schmidt}, {Skottfelt},
  {Southworth}, {Starkey}, {Surdej}, {Wertz}, {Zarucki}, {MiNDSTEp Consortium},
  {Gaudi}, {Pogge}, {DePoy}, \& {{$\mu$}FUN Collaboration}}]{CalchiNovati2015a}
{Calchi Novati}, S., {Gould}, A., {Udalski}, A., {et~al.} 2015{\natexlab{a}},
  \apj, 804, 20

\bibitem[{{Calchi Novati} {et~al.}(2015{\natexlab{b}}){Calchi Novati}, {Gould},
  {Yee}, {Beichman}, {Bryden}, {Carey}, {Fausnaugh}, {Gaudi}, {Henderson},
  {Pogge}, {Shvartzvald}, {Wibking}, {Zhu}, {Spitzer Team}, {Udalski},
  {Poleski}, {Pawlak}, {Szyma{\'n}ski}, {Skowron}, {Mr{\'o}z}, {Koz{\l}owski},
  {Wyrzykowski}, {Pietrukowicz}, {Pietrzy{\'n}ski}, {Soszy{\'n}ski}, {Ulaczyk},
  \& {OGLE Group}}]{CalchiNovati2015b}
{Calchi Novati}, S., {Gould}, A., {Yee}, J.~C., {et~al.} 2015{\natexlab{b}},
  \apj, 814, 92

\bibitem[{{Calchi Novati}(2018)}]{CalchiNovati2017}
{Calchi Novati}, S. e.~a. 2018, AAS, submitted

\bibitem[{{Carson} {et~al.}(2013){Carson}, {Thalmann}, {Janson}, {Kozakis},
  {Bonnefoy}, {Biller}, {Schlieder}, {Currie}, {McElwain}, {Goto}, {Henning},
  {Brandner}, {Feldt}, {Kandori}, {Kuzuhara}, {Stevens}, {Wong}, {Gainey},
  {Fukagawa}, {Kuwada}, {Brandt}, {Kwon}, {Abe}, {Egner}, {Grady}, {Guyon},
  {Hashimoto}, {Hayano}, {Hayashi}, {Hayashi}, {Hodapp}, {Ishii}, {Iye},
  {Knapp}, {Kudo}, {Kusakabe}, {Matsuo}, {Miyama}, {Morino}, {Moro-Martin},
  {Nishimura}, {Pyo}, {Serabyn}, {Suto}, {Suzuki}, {Takami}, {Takato},
  {Terada}, {Tomono}, {Turner}, {Watanabe}, {Wisniewski}, {Yamada}, {Takami},
  {Usuda}, \& {Tamura}}]{Carson2013}
{Carson}, J., {Thalmann}, C., {Janson}, M., {et~al.} 2013, \apjl, 763, L32

\bibitem[{{Chabrier}(2003)}]{Chabrier2003}
{Chabrier}, G. 2003, \pasp, 115, 763

\bibitem[{{Choi} {et~al.}(2013){Choi}, {Han}, {Udalski}, {Sumi}, {Gaudi},
  {Gould}, {Bennett}, {Dominik}, {Beaulieu}, {Tsapras}, {Bozza}, {Abe}, {Bond},
  {Botzler}, {Chote}, {Freeman}, {Fukui}, {Furusawa}, {Itow}, {Ling}, {Masuda},
  {Matsubara}, {Miyake}, {Muraki}, {Ohnishi}, {Rattenbury}, {Saito},
  {Sullivan}, {Suzuki}, {Sweatman}, {Suzuki}, {Takino}, {Tristram}, {Wada},
  {Yock}, {MOA Collaboration}, {Szyma{\'n}ski}, {Kubiak}, {Pietrzy{\'n}ski},
  {Soszy{\'n}ski}, {Skowron}, {Koz{\l}owski}, {Poleski}, {Ulaczyk},
  {Wyrzykowski}, {Pietrukowicz}, {OGLE Collaboration}, {Almeida}, {DePoy},
  {Dong}, {Gorbikov}, {Jablonski}, {Henderson}, {Hwang}, {Janczak}, {Jung},
  {Kaspi}, {Lee}, {Malamud}, {Maoz}, {McGregor}, {Mu{\~n}oz}, {Park}, {Park},
  {Pogge}, {Shvartzvald}, {Shin}, {Yee}, {{$\mu$}FUN Collaboration}, {Alsubai},
  {Browne}, {Burgdorf}, {Calchi Novati}, {Dodds}, {Fang}, {Finet}, {Glitrup},
  {Grundahl}, {Gu}, {Hardis}, {Harps{\o}e}, {Hinse}, {Hornstrup},
  {Hundertmark}, {Jessen-Hansen}, {Jrgensen}, {Kains}, {Kerins}, {Liebig},
  {Lund}, {Lundkvist}, {Maier}, {Mancini}, {Mathiasen}, {Penny}, {Rahvar},
  {Ricci}, {Scarpetta}, {Skottfelt}, {Snodgrass}, {Southworth}, {Surdej},
  {Tregloan-Reed}, {Wambsganss}, {Wertz}, {Zimmer}, {MiNDSTEp Consortium},
  {Albrow}, {Bachelet}, {Batista}, {Brillant}, {Cassan}, {Cole}, {Coutures},
  {Dieters}, {Dominis Prester}, {Donatowicz}, {Fouqu{\'e}}, {Greenhill},
  {Kubas}, {Marquette}, {Menzies}, {Sahu}, {Zub}, {PLANET Collaboration},
  {Bramich}, {Horne}, {Steele}, {Street}, \& {RoboNet
  Collaboration}}]{Choi2013}
{Choi}, J.-Y., {Han}, C., {Udalski}, A., {et~al.} 2013, \apj, 768, 129

\bibitem[{{Chung} {et~al.}(2017){Chung}, {Zhu}, {Udalski}, {Lee}, {Ryu},
  {Jung}, {Shin}, {Yee}, {Hwang}, {Gould}, {and}, {Albrow}, {Cha}, {Han},
  {Kim}, {Kim}, {Kim}, {Kim}, {Lee}, {Park}, {Pogge}, {KMTNet Collaboration},
  {Poleski}, {Mr{\'o}z}, {Pietrukowicz}, {Skowron}, {Szyma{\'n}ski},
  {Soszy{\'n}ski}, {Koz{\l}owski}, {Ulaczyk}, {Pawlak}, {OGLE Collaboration},
  {Beichman}, {Bryden}, {Calchi Novati}, {Carey}, {Fausnaugh}, {Gaudi},
  {Henderson}, {Shvartzvald}, {Wibking}, \& {The Spitzer Team}}]{Chung2017}
{Chung}, S.-J., {Zhu}, W., {Udalski}, A., {et~al.} 2017, \apj, 838, 154

\bibitem[{{Dong} {et~al.}(2009){Dong}, {Gould}, {Udalski}, {Anderson},
  {Christie}, {Gaudi}, {OGLE Collaboration}, {Jaroszy{\'n}ski}, {Kubiak},
  {Szyma{\'n}ski}, {Pietrzy{\'n}ski}, {Soszy{\'n}ski}, {Szewczyk}, {Ulaczyk},
  {Wyrzykowski}, {{$\mu$}FUN Collaboration}, {DePoy}, {Fox}, {Gal-Yam}, {Han},
  {L{\'e}pine}, {McCormick}, {Ofek}, {Park}, {Pogge}, {MOA Collaboration},
  {Abe}, {Bennett}, {Bond}, {Britton}, {Gilmore}, {Hearnshaw}, {Itow},
  {Kamiya}, {Kilmartin}, {Korpela}, {Masuda}, {Matsubara}, {Motomura},
  {Muraki}, {Nakamura}, {Ohnishi}, {Okada}, {Rattenbury}, {Saito}, {Sako},
  {Sasaki}, {Sullivan}, {Sumi}, {Tristram}, {Yanagisawa}, {Yock}, {Yoshoika},
  {PLANET/RoboNet Collaborations}, {Albrow}, {Beaulieu}, {Brillant}, {Calitz},
  {Cassan}, {Cook}, {Coutures}, {Dieters}, {Dominis Prester}, {Donatowicz},
  {Fouqu{\'e}}, {Greenhill}, {Hill}, {Hoffman}, {Horne}, {J{\o}rgensen},
  {Kane}, {Kubas}, {Marquette}, {Martin}, {Meintjes}, {Menzies}, {Pollard},
  {Sahu}, {Vinter}, {Wambsganss}, {Williams}, {Bode}, {Bramich}, {Burgdorf},
  {Snodgrass}, {Steele}, {Doublier}, \& {Foellmi}}]{Dong2009}
{Dong}, S., {Gould}, A., {Udalski}, A., {et~al.} 2009, \apj, 695, 970

\bibitem[{{Dupuy} \& {Liu}(2017)}]{Dupuy2017}
{Dupuy}, T.~J., \& {Liu}, M.~C. 2017, \apjs, 231, 15

\bibitem[{{Fazio} {et~al.}(2004){Fazio}, {Hora}, {Allen}, {Ashby}, {Barmby},
  {Deutsch}, {Huang}, {Kleiner}, {Marengo}, {Megeath}, {Melnick}, {Pahre},
  {Patten}, {Polizotti}, {Smith}, {Taylor}, {Wang}, {Willner}, {Hoffmann},
  {Pipher}, {Forrest}, {McMurty}, {McCreight}, {McKelvey}, {McMurray}, {Koch},
  {Moseley}, {Arendt}, {Mentzell}, {Marx}, {Losch}, {Mayman}, {Eichhorn},
  {Krebs}, {Jhabvala}, {Gezari}, {Fixsen}, {Flores}, {Shakoorzadeh}, {Jungo},
  {Hakun}, {Workman}, {Karpati}, {Kichak}, {Whitley}, {Mann}, {Tollestrup},
  {Eisenhardt}, {Stern}, {Gorjian}, {Bhattacharya}, {Carey}, {Nelson},
  {Glaccum}, {Lacy}, {Lowrance}, {Laine}, {Reach}, {Stauffer}, {Surace},
  {Wilson}, {Wright}, {Hoffman}, {Domingo}, \& {Cohen}}]{Fazio2004}
{Fazio}, G.~G., {Hora}, J.~L., {Allen}, L.~E., {et~al.} 2004, \apjs, 154, 10

\bibitem[{{Foreman-Mackey} {et~al.}(2013){Foreman-Mackey}, {Hogg}, {Lang}, \&
  {Goodman}}]{ForemanMackey2013}
{Foreman-Mackey}, D., {Hogg}, D.~W., {Lang}, D., \& {Goodman}, J. 2013, \pasp,
  125, 306

\bibitem[{{Furusawa} {et~al.}(2013){Furusawa}, {Udalski}, {Sumi}, {Bennett},
  {Bond}, {Gould}, {J{\o}rgensen}, {Snodgrass}, {Dominis Prester}, {Albrow},
  {Abe}, {Botzler}, {Chote}, {Freeman}, {Fukui}, {Harris}, {Itow}, {Ling},
  {Masuda}, {Matsubara}, {Miyake}, {Muraki}, {Ohnishi}, {Rattenbury}, {Saito},
  {Sullivan}, {Suzuki}, {Sweatman}, {Tristram}, {Wada}, {Yock}, {MOA
  Collaboration}, {Szyma{\'n}ski}, {Soszy{\'n}ski}, {Kubiak}, {Poleski},
  {Ulaczyk}, {Pietrzy{\'n}ski}, {Wyrzykowski}, {OGLE Collaboration}, {Choi},
  {Christie}, {DePoy}, {Dong}, {Drummond}, {Gaudi}, {Han}, {Hung}, {Hwang},
  {Lee}, {McCormick}, {Moorhouse}, {Natusch}, {Nola}, {Ofek}, {Pogge}, {Shin},
  {Skowron}, {Thornley}, {Yee}, {{$\mu$}FUN Collaboration}, {Alsubai}, {Bozza},
  {Browne}, {Burgdorf}, {Calchi Novati}, {Dodds}, {Dominik}, {Finet}, {Gerner},
  {Hardis}, {Harps{\o}e}, {Hinse}, {Hundertmark}, {Kains}, {Kerins}, {Liebig},
  {Mancini}, {Mathiasen}, {Penny}, {Proft}, {Rahvar}, {Ricci}, {Scarpetta},
  {Sch{\"a}fer}, {Sch{\"o}nebeck}, {Southworth}, {Surdej}, {Wambsganss},
  {MiNDSTEp Consortium}, {Street}, {Bramich}, {Steele}, {Tsapras}, {RoboNet
  Collaboration}, {Horne}, {Donatowicz}, {Sahu}, {Bachelet}, {Batista},
  {Beatty}, {Beaulieu}, {Bennett}, {Black}, {Bowens-Rubin}, {Brillant},
  {Caldwell}, {Cassan}, {Cole}, {Corrales}, {Coutures}, {Dieters},
  {Fouqu{\'e}}, {Greenhill}, {Henderson}, {Kubas}, {Marquette}, {Martin},
  {Menzies}, {Shappee}, {Williams}, {Wouters}, {van Saders}, {Zellem}, {Zub},
  \& {PLANET Collaboration}}]{Furusawa2013}
{Furusawa}, K., {Udalski}, A., {Sumi}, T., {et~al.} 2013, \apj, 779, 91

\bibitem[{{Gagn{\'e}} {et~al.}(2017){Gagn{\'e}}, {Faherty}, {Mamajek}, {Malo},
  {Doyon}, {Filippazzo}, {Weinberger}, {Donaldson}, {L{\'e}pine},
  {Lafreni{\`e}re}, {Artigau}, {Burgasser}, {Looper}, {Boucher}, {Beletsky},
  {Camnasio}, {Brunette}, \& {Arboit}}]{Gagne2017}
{Gagn{\'e}}, J., {Faherty}, J.~K., {Mamajek}, E.~E., {et~al.} 2017, \apjs, 228,
  18

\bibitem[{{Gould}(1992)}]{Gould1992}
{Gould}, A. 1992, \apj, 392, 442

\bibitem[{{Gould}(1994)}]{Gould1994}
---. 1994, \apjl, 421, L75

\bibitem[{{Gould}(2000)}]{Gould2000}
---. 2000, \apj, 542, 785

\bibitem[{{Gould}(2004)}]{Gould2004}
---. 2004, \apj, 606, 319

\bibitem[{{Gould}(2008)}]{Gould2008}
---. 2008, \apj, 681, 1593

\bibitem[{{Gould}(2016)}]{Gould2016}
---. 2016, Journal of Korean Astronomical Society, 49, 123

\bibitem[{{Gould} {et~al.}(2009){Gould}, {Udalski}, {Monard}, {Horne}, {Dong},
  {Miyake}, {Sahu}, {Bennett}, {Wyrzykowski}, {Soszy{\'n}ski}, {Szyma{\'n}ski},
  {Kubiak}, {Pietrzy{\'n}ski}, {Szewczyk}, {Ulaczyk}, {OGLE Collaboration},
  {Allen}, {Christie}, {DePoy}, {Gaudi}, {Han}, {Lee}, {McCormick}, {Natusch},
  {Park}, {Pogge}, {{$\mu$}FUN Collaboration}, {Allan}, {Bode}, {Bramich},
  {Burgdorf}, {Dominik}, {Fraser}, {Kerins}, {Mottram}, {Snodgrass}, {Steele},
  {Street}, {Tsapras}, {RoboNet Collaboration}, {Abe}, {Bond}, {Botzler},
  {Fukui}, {Furusawa}, {Hearnshaw}, {Itow}, {Kamiya}, {Kilmartin}, {Korpela},
  {Lin}, {Ling}, {Masuda}, {Matsubara}, {Muraki}, {Nagaya}, {Ohnishi},
  {Okumura}, {Perrott}, {Rattenbury}, {Saito}, {Sako}, {Skuljan}, {Sullivan},
  {Sumi}, {Sweatman}, {Tristram}, {Yock}, {MOA Collaboration}, {Albrow},
  {Beaulieu}, {Coutures}, {Calitz}, {Caldwell}, {Fouque}, {Martin}, {Williams},
  \& {PLANET Collaboration}}]{Gould2009}
{Gould}, A., {Udalski}, A., {Monard}, B., {et~al.} 2009, \apjl, 698, L147

\bibitem[{{Han} \& {Chang}(2003)}]{Han2003b}
{Han}, C., \& {Chang}, H.-Y. 2003, \mnras, 338, 637

\bibitem[{{Han} \& {Gould}(2003)}]{Han2003a}
{Han}, C., \& {Gould}, A. 2003, \apj, 592, 172

\bibitem[{{Han} {et~al.}(2017{\natexlab{a}}){Han}, {Udalski}, {Sumi}, {Gould},
  {Albrow}, {Chung}, {Jung}, {Ryu}, {Shin}, {Yee}, {Zhu}, {Cha}, {Kim}, {Kim},
  {Lee}, {Lee}, {Park}, {The KMTNet Collaboration}, {Soszy{\'n}ski},
  {Mr{\'o}z}, {Pietrukowicz}, {Szyma{\'n}ski}, {Skowron}, {Poleski},
  {Koz{\l}owski}, {Ulaczyk}, {Pawlak}, {The OGLE Collaboration}, {Abe},
  {Asakura}, {Bennett}, {Bond}, {Bhattacharya}, {Donachie}, {Freeman}, {Fukui},
  {Hirao}, {Itow}, {Koshimoto}, {Li}, {Ling}, {Masuda}, {Matsubara}, {Muraki},
  {Nagakane}, {Ohnishi}, {Oyokawa}, {Rattenbury}, {Saito}, {Sharan},
  {Sullivan}, {Suzuki}, {Tristram}, {Yamada}, {Yamada}, {Yonehara}, {Barry}, \&
  {The MOA Collaboration}}]{Han2017b}
{Han}, C., {Udalski}, A., {Sumi}, T., {et~al.} 2017{\natexlab{a}}, \apj, 843,
  59

\bibitem[{{Han} {et~al.}(2017{\natexlab{b}}){Han}, {Udalski}, {Bozza},
  {Szyma{\'n}ski}, {Soszy{\'n}ski}, {Skowron}, {Mr{\'o}z}, {Poleski},
  {Pietrukowicz}, {Koz{\l}owski}, {Ulaczyk}, {Wyrzykowski}, {The OGLE
  Collaboration}, {Calchi Novati}, {D'Ago}, {Dominik}, {Hundertmark},
  {Jorgensen}, {Scarpetta}, \& {The MiNDSTEp Consortium}}]{Han2017a}
{Han}, C., {Udalski}, A., {Bozza}, V., {et~al.} 2017{\natexlab{b}}, \apj, 843,
  87

\bibitem[{{Jeffries}(2012)}]{Jeffries2012}
{Jeffries}, R.~D. 2012, in EAS Publications Series, Vol.~57, EAS Publications
  Series, ed. C.~{Reyl{\'e}}, C.~{Charbonnel}, \& M.~{Schultheis}, 45--89

\bibitem[{{Kervella} {et~al.}(2004){Kervella}, {Th{\'e}venin}, {Di Folco}, \&
  {S{\'e}gransan}}]{Kervella2004}
{Kervella}, P., {Th{\'e}venin}, F., {Di Folco}, E., \& {S{\'e}gransan}, D.
  2004, \aap, 426, 297

\bibitem[{{Kim} {et~al.}(2017){Kim}, {Kim}, {Hwang}, {Albrow}, {Chung},
  {Gould}, {Han}, {Jung}, {Ryu}, {Shin}, {Yee}, {Zhu}, {Cha}, {Kim}, {Lee},
  {Lee}, {Park}, \& {Pogge}}]{Kim2017}
{Kim}, D.-J., {Kim}, H.-W., {Hwang}, K.-H., {et~al.} 2017, ArXiv e-prints,
  arXiv:1703.06883

\bibitem[{{Kim}(2018 in preparation)}]{Kim2018}
{Kim}, H.-W. e.~a. 2018 in preparation

\bibitem[{{Kim} {et~al.}(2016){Kim}, {Lee}, {Park}, {Kim}, {Cha}, {Lee}, {Han},
  {Chun}, \& {Yuk}}]{Kim2016}
{Kim}, S.-L., {Lee}, C.-U., {Park}, B.-G., {et~al.} 2016, Journal of Korean
  Astronomical Society, 49, 37

\bibitem[{{Kroupa}(2002)}]{Kroupa2002}
{Kroupa}, P. 2002, Science, 295, 82

\bibitem[{{Lovis} \& {Mayor}(2007)}]{Lovis2007}
{Lovis}, C., \& {Mayor}, M. 2007, \aap, 472, 657

\bibitem[{{Luhman}(2013)}]{Luhman2013}
{Luhman}, K.~L. 2013, \apjl, 767, L1

\bibitem[{{McDougall} \& {Albrow}(2016)}]{McDougall2016}
{McDougall}, A., \& {Albrow}, M.~D. 2016, \mnras, 456, 565

\bibitem[{{Mordasini} {et~al.}(2009){Mordasini}, {Alibert}, \&
  {Benz}}]{Mordasini2009}
{Mordasini}, C., {Alibert}, Y., \& {Benz}, W. 2009, \aap, 501, 1139

\bibitem[{{Mr{\'o}z} {et~al.}(2017){Mr{\'o}z}, {Udalski}, {Skowron}, {Poleski},
  {Koz{\l}owski}, {Szyma{\'n}ski}, {Soszy{\'n}ski}, {Wyrzykowski},
  {Pietrukowicz}, {Ulaczyk}, {Skowron}, \& {Pawlak}}]{Mroz2017}
{Mr{\'o}z}, P., {Udalski}, A., {Skowron}, J., {et~al.} 2017, \nat, 548, 183

\bibitem[{{Mroz} {et~al.}(2018){Mroz}, {Ryu}, {Skowron}, {Udalski}, {Gould},
  {Szymanski}, {Soszynski}, {Poleski}, {Pietrukowicz}, {Kozlowski}, {Pawlak},
  {Ulaczyk}, {Albrow}, {Chung}, {Jung}, {Han}, {Hwang}, {Shin}, {Yee}, {Zhu},
  {Cha}, {Kim}, {Kim}, {Kim}, {Lee}, {Lee}, {Lee}, {Park}, \&
  {Pogge}}]{Mroz2018}
{Mroz}, P., {Ryu}, Y.-H., {Skowron}, J., {et~al.} 2018, AJ, in press,
  arXiv:1712.01042

\bibitem[{{Nataf} {et~al.}(2013){Nataf}, {Gould}, {Fouqu{\'e}}, {Gonzalez},
  {Johnson}, {Skowron}, {Udalski}, {Szyma{\'n}ski}, {Kubiak},
  {Pietrzy{\'n}ski}, {Soszy{\'n}ski}, {Ulaczyk}, {Wyrzykowski}, \&
  {Poleski}}]{Nataf2013}
{Nataf}, D.~M., {Gould}, A., {Fouqu{\'e}}, P., {et~al.} 2013, \apj, 769, 88

\bibitem[{{Park} {et~al.}(2004){Park}, {DePoy}, {Gaudi}, {Gould}, {Han},
  {Pogge}, {muFun Collaboration}, {Abe}, {Bennett}, {Bond}, {Eguchi}, {Furuta},
  {Hearnshaw}, {Kamiya}, {Kilmartin}, {Kurata}, {Masuda}, {Matsubara},
  {Muraki}, {Noda}, {Okajima}, {Rattenbury}, {Sako}, {Sekiguchi}, {Sullivan},
  {Sumi}, {Tristram}, {Yanagisawa}, {Yock}, \& {MOA Collaboration}}]{Park2004}
{Park}, B.-G., {DePoy}, D.~L., {Gaudi}, B.~S., {et~al.} 2004, \apj, 609, 166

\bibitem[{{Park} {et~al.}(2015){Park}, {Udalski}, {Han}, {Poleski}, {Skowron},
  {Koz{\l}owski}, {Wyrzykowski}, {Szyma{\'n}ski}, {Pietrukowicz},
  {Pietrzy{\'n}ski}, {Soszy{\'n}ski}, {Ulaczyk}, \& {OGLE
  Collaboration}}]{Park2015}
{Park}, H., {Udalski}, A., {Han}, C., {et~al.} 2015, \apj, 805, 117

\bibitem[{{Pejcha} \& {Heyrovsk{\'y}}(2009)}]{Pejcha2009}
{Pejcha}, O., \& {Heyrovsk{\'y}}, D. 2009, \apj, 690, 1772

\bibitem[{{Penny} {et~al.}(2016){Penny}, {Henderson}, \& {Clanton}}]{Penny2016}
{Penny}, M.~T., {Henderson}, C.~B., \& {Clanton}, C. 2016, \apj, 830, 150

\bibitem[{{Refsdal}(1966)}]{Refsdal1966}
{Refsdal}, S. 1966, \mnras, 134, 315

\bibitem[{{Saito} {et~al.}(2012){Saito}, {Hempel}, {Minniti}, {Lucas},
  {Rejkuba}, {Toledo}, {Gonzalez}, {Alonso-Garc{\'{\i}}a}, {Irwin},
  {Gonzalez-Solares}, {Hodgkin}, {Lewis}, {Cross}, {Ivanov}, {Kerins},
  {Emerson}, {Soto}, {Am{\^o}res}, {Gurovich}, {D{\'e}k{\'a}ny}, {Angeloni},
  {Beamin}, {Catelan}, {Padilla}, {Zoccali}, {Pietrukowicz}, {Moni Bidin},
  {Mauro}, {Geisler}, {Folkes}, {Sale}, {Borissova}, {Kurtev}, {Ahumada},
  {Alonso}, {Adamson}, {Arias}, {Bandyopadhyay}, {Barb{\'a}}, {Barbuy},
  {Baume}, {Bedin}, {Bellini}, {Benjamin}, {Bica}, {Bonatto}, {Bronfman},
  {Carraro}, {Chen{\`e}}, {Clari{\'a}}, {Clarke}, {Contreras}, {Corvill{\'o}n},
  {de Grijs}, {Dias}, {Drew}, {Fari{\~n}a}, {Feinstein},
  {Fern{\'a}ndez-Laj{\'u}s}, {Gamen}, {Gieren}, {Goldman},
  {Gonz{\'a}lez-Fern{\'a}ndez}, {Grand}, {Gunthardt}, {Hambly}, {Hanson},
  {He{\l}miniak}, {Hoare}, {Huckvale}, {Jord{\'a}n}, {Kinemuchi}, {Longmore},
  {L{\'o}pez-Corredoira}, {Maccarone}, {Majaess}, {Mart{\'{\i}}n}, {Masetti},
  {Mennickent}, {Mirabel}, {Monaco}, {Morelli}, {Motta}, {Palma}, {Parisi},
  {Parker}, {Pe{\~n}aloza}, {Pietrzy{\'n}ski}, {Pignata}, {Popescu}, {Read},
  {Rojas}, {Roman-Lopes}, {Ruiz}, {Saviane}, {Schreiber}, {Schr{\"o}der},
  {Sharma}, {Smith}, {Sodr{\'e}}, {Stead}, {Stephens}, {Tamura}, {Tappert},
  {Thompson}, {Valenti}, {Vanzi}, {Walton}, {Weidmann}, \&
  {Zijlstra}}]{Saito2012}
{Saito}, R.~K., {Hempel}, M., {Minniti}, D., {et~al.} 2012, \aap, 537, A107

\bibitem[{{Schechter} {et~al.}(1993){Schechter}, {Mateo}, \&
  {Saha}}]{Schechter1993}
{Schechter}, P.~L., {Mateo}, M., \& {Saha}, A. 1993, \pasp, 105, 1342

\bibitem[{{Schlaufman}(2018)}]{Schlaufman2018}
{Schlaufman}, K.~C. 2018, \apj, 853, 37

\bibitem[{{Shin} {et~al.}(2017){Shin}, {Udalski}, {Yee}, {Calchi Novati},
  {Han}, {Skowron}, {Mr{\'o}z}, {Soszy{\'n}ski}, {Poleski}, {Szyma{\'n}ski},
  {Koz{\l}owski}, {Pietrukowicz}, {Ulaczyk}, {Pawlak}, {OGLE Collaboration},
  {Albrow}, {Gould}, {Chung}, {Hwang}, {Jung}, {Ryu}, {Zhu}, {Cha}, {Kim},
  {Kim}, {Kim}, {Lee}, {Lee}, {Park}, {Pogge}, {KMTNet Group}, {Beichman},
  {Bryden}, {Carey}, {Gaudi}, {Henderson}, {Shvartzvald}, \& {Spitzer
  Team}}]{Shin2017}
{Shin}, I.-G., {Udalski}, A., {Yee}, J.~C., {et~al.} 2017, \aj, 154, 176

\bibitem[{{Shin} {et~al.}(2018){Shin}, {Udalski}, {Yee}, {Calchi Novati},
  {Christie}, {Poleski}, {Mr{\'o}z}, {Skowron}, {Szyma{\'n}ski},
  {Soszy{\'n}ski}, {Pietrukowicz}, {Koz{\l}owski}, {Ulaczyk}, {Pawlak},
  {Natusch}, {Pogge}, {Gould}, {Han}, {Albrow}, {Chung}, {Hwang}, {Ryu},
  {Jung}, {Zhu}, {Lee}, {Cha}, {Kim}, {Kim}, {Kim}, {Lee}, {Lee}, {Park},
  {Beichman}, {Bryden}, {Carey}, {Gaudi}, {Henderson}, \&
  {Shvartzvald}}]{Shin2018}
---. 2018, ArXiv e-prints, arXiv:1801.00169

\bibitem[{{Shvartzvald} {et~al.}(2017){Shvartzvald}, {Yee}, {Calchi Novati},
  {Gould}, {Lee}, {Beichman}, {Bryden}, {Carey}, {Gaudi}, {Henderson}, {Zhu},
  {Spitzer team}, {Albrow}, {Cha}, {Chung}, {Han}, {Hwang}, {Jung}, {Kim},
  {Kim}, {Kim}, {Lee}, {Park}, {Pogge}, {Ryu}, {Shin}, \& {KMTNet
  group}}]{Shvartzvald2017}
{Shvartzvald}, Y., {Yee}, J.~C., {Calchi Novati}, S., {et~al.} 2017, \apjl,
  840, L3

\bibitem[{{Udalski} {et~al.}(2015){Udalski}, {Szyma{\'n}ski}, \&
  {Szyma{\'n}ski}}]{Udalski2015}
{Udalski}, A., {Szyma{\'n}ski}, M.~K., \& {Szyma{\'n}ski}, G. 2015, \actaa, 65,
  1

\bibitem[{{Yee} {et~al.}(2009){Yee}, {Udalski}, {Sumi}, {Dong}, {Koz{\l}owski},
  {Bird}, {Cole}, {Higgins}, {McCormick}, {Monard}, {Polishook}, {Shporer},
  {Spector}, {Szyma{\'n}ski}, {Kubiak}, {Pietrzy{\'n}ski}, {Soszy{\'n}ski},
  {Szewczyk}, {Ulaczyk}, {Wyrzykowski}, {Poleski}, {OGLE Collaboration},
  {Allen}, {Bos}, {Christie}, {DePoy}, {Eastman}, {Gaudi}, {Gould}, {Han},
  {Kaspi}, {Lee}, {Mallia}, {Maury}, {Maoz}, {Natusch}, {Park}, {Pogge},
  {Santallo}, {{$\mu$}FUN Collaboration}, {Abe}, {Bond}, {Fukui}, {Furusawa},
  {Hearnshaw}, {Hosaka}, {Itow}, {Kamiya}, {Korpela}, {Kilmartin}, {Lin},
  {Ling}, {Makita}, {Masuda}, {Matsubara}, {Miyake}, {Muraki}, {Nagaya},
  {Nishimoto}, {Ohnishi}, {Perrott}, {Rattenbury}, {Sako}, {Saito}, {Skuljan},
  {Sullivan}, {Sweatman}, {Tristram}, {Yock}, {MOA Collaboration}, {Albrow},
  {Batista}, {Fouqu{\'e}}, {Beaulieu}, {Bennett}, {Cassan}, {Comparat},
  {Coutures}, {Dieters}, {Greenhill}, {Horne}, {Kains}, {Kubas}, {Martin},
  {Menzies}, {Wambsganss}, {Williams}, {Zub}, \& {PLANET
  Collaboration}}]{Yee2009}
{Yee}, J.~C., {Udalski}, A., {Sumi}, T., {et~al.} 2009, \apj, 703, 2082

\bibitem[{{Yee} {et~al.}(2015){Yee}, {Gould}, {Beichman}, {Calchi Novati},
  {Carey}, {Gaudi}, {Henderson}, {Nataf}, {Penny}, {Shvartzvald}, \&
  {Zhu}}]{Yee2015}
{Yee}, J.~C., {Gould}, A., {Beichman}, C., {et~al.} 2015, \apj, 810, 155

\bibitem[{{Zhu} {et~al.}(2016){Zhu}, {Calchi Novati}, {Gould}, {Udalski},
  {Han}, {Shvartzvald}, {Ranc}, {J{\o}rgensen}, {Poleski}, {Bozza}, {Beichman},
  {Bryden}, {Carey}, {Gaudi}, {Henderson}, {Pogge}, {Porritt}, {Wibking},
  {Yee}, {SPITZER Team}, {Pawlak}, {Szyma{\'n}ski}, {Skowron}, {Mr{\'o}z},
  {Koz{\l}owski}, {Wyrzykowski}, {Pietrukowicz}, {Pietrzy{\'n}ski},
  {Soszy{\'n}ski}, {Ulaczyk}, {OGLE Group}, {Choi}, {Park}, {Jung}, {Shin},
  {Albrow}, {Park}, {Kim}, {Lee}, {Cha}, {Kim}, {Lee}, {KMTNET Group},
  {Friedmann}, {Kaspi}, {Maoz}, {WISE Group}, {Hundertmark}, {Street},
  {Tsapras}, {Bramich}, {Cassan}, {Dominik}, {Bachelet}, {Dong}, {Figuera
  Jaimes}, {Horne}, {Mao}, {Menzies}, {Schmidt}, {Snodgrass}, {Steele},
  {Wambsganss}, {RoboNeT Team}, {Skottfelt}, {Andersen}, {Burgdorf}, {Ciceri},
  {D'Ago}, {Evans}, {Gu}, {Hinse}, {Kerins}, {Korhonen}, {Kuffmeier},
  {Mancini}, {Peixinho}, {Popovas}, {Rabus}, {Rahvar}, {Tronsgaard},
  {Scarpetta}, {Southworth}, {Surdej}, {von Essen}, {Wang}, {Wertz}, \&
  {MiNDSTEP Group}}]{Zhu2016}
{Zhu}, W., {Calchi Novati}, S., {Gould}, A., {et~al.} 2016, \apj, 825, 60

\end{thebibliography}

\newpage

\newpage

\begin{deluxetable}{| l | cc | cc | cc | cc | }[h]
\rotate
\tabletypesize{\scriptsize}
\tablecaption{Microlensing parameters, physical parameters and relative probabilities for the combined Spitzer and ground-based photometry for the A (green) solutions. 
The $+/-$ geometry refers
to the signs of $u_{\rm 0,Earth}$ / $u_{\rm 0,{\it Spitzer}}$. The left/right column for each geometry gives the parameters
for the models without/with lens orbital motion.}
\tablehead{
  & \multicolumn{2}{c|}{$+/+$} & \multicolumn{2}{c|}{$+/-$} & \multicolumn{2}{c|}{$-/+$} & \multicolumn{2}{c|}{$-/-$} 
}
\startdata
$\log_{\rm 10}s$ & ${-0.1852}_{-0.0022}^{+0.0024}$ & ${-0.178}_{-0.0027}^{+0.0024}$ & ${-0.1849}_{-0.0024}^{+0.0023}$ & ${-0.
1777}_{-0.0029}^{+0.0024}$ & ${-0.1845}_{-0.0024}^{+0.0024}$ & ${-0.1770}_{-0.0029}^{+0.0022}$ & ${-0.1844}_{-0.0023}^{+0.0023}$ & ${-0.1773}_{-0.0027}^{+0.0023}$ \\
$\log_{\rm 10}q$ & ${-0.150}_{-0.012}^{+0.012}$ & ${-0.118}_{-0.014}^{+0.015}$ & ${-0.150}_{-0.013}^{+0.012}$ & ${-0.118}_{
-0.016}^{+0.016}$ & ${-0.149}_{-0.012}^{+0.012}$ & ${-0.119}_{-0.014}^{+0.015}$ & ${-0.149}_{-0.013}^{+0.013}$ & ${-
0.119}_{-0.016}^{+0.015}$ \\
$log_{\rm 10}\rho$ & ${-1.599}_{-0.004}^{+0.004}$ & ${-1.550}_{-0.018}^{+0.010}$ & ${-1.599}_{-0.004}^{+0.005}$ & ${-1.549}
_{-0.018}^{+0.010}$ & ${-1.598}_{-0.005}^{+0.004}$ & ${-1.547}_{-0.016}^{+0.009}$ & ${-1.598}_{-0.004}^{+0.004}$ 
& ${-1.548}_{-0.016}^{+0.009}$ \\
$u_{\rm 0}$ & ${1.181}_{-0.011}^{+0.011}$ & ${1.154}_{-0.011}^{+0.011}$ & ${1.180}_{-0.011}^{+0.011}$ & ${1.154}_{-0.010}^{+0.
011}$ & ${-1.179}_{-0.012}^{+0.012}$ & ${-1.152}_{-0.011}^{+0.010}$ & ${-1.178}_{-0.011}^{+0.011}$ & ${-1.154}_{-0.011}^{+0.010}$ \\
$\alpha$ & ${3.177}_{-0.013}^{+0.014}$ & ${3.137}_{-0.016}^{+0.016}$ & ${3.177}_{-0.013}^{+0.013}$ & ${3.137}_{-0.019}^{+0
.018}$ & ${3.107}_{-0.014}^{+0.013}$ & ${3.144}_{-0.015}^{+0.016}$ & ${3.106}_{-0.014}^{+0.015}$ & ${3.144}_{-0.018}^{+
0.017}$ \\
$t_{\rm 0}$ & ${7583.54}_{-0.06}^{+0.05}$ & ${7583.47}_{-0.06}^{+0.06}$ & ${7583.54}_{-0.06}^{+0.06}$ & ${7583.46}_{-0.06}^{+0.06}$ & ${7583.54}_{-0.05}^{+0.05}$ & ${7583.47}_{-0.06}^{+0.06}$ & ${7583.53}_{-0.05}^{+0.05}$ & ${7583.47}_{-0.06}^{+0.06}$ \\
$t_{\rm E}$ & ${8.84}_{-0.06}^{+0.07}$ & ${8.68}_{-0.07}^{+0.08}$ & ${8.84}_{-0.06}^{+0.06}$ & ${8.67}_{-0.07}^{+0.08}$ & ${8.81}_{-0.06}^{+0.07}$ & ${8.65}_{-0.07}^{+0.08}$ & ${8.82}_{-0.07}^{+0.07}$ & ${8.65}_{-0.07}^{+0.08}$ \\
$\pi_{\rm E,E}$ & ${0.482}_{-0.018}^{+0.018}$ & ${0.481}_{-0.019}^{+0.017}$ & ${0.481}_{-0.018}^{+0.016}$ & ${0.479}_{-0.019}^{+0.017}$ & ${0.413}_{-0.018}^{+0.016}$ & ${0.413}_{-0.019}^{+0.017}$ & ${0.411}_{-0.017}^{+0.016}$ & ${0.410}_{
-0.019}^{+0.018}$ \\
$\pi_{\rm E,N}$ & ${-0.86}_{-0.10}^{+0.09}$ & ${-0.89}_{-0.09}^{+0.10}$ & ${-0.86}_{-0.10}^{+0.10}$ & ${-0.86}_{-0
.11}^{+0.11}$ & ${0.88}_{-0.08}^{+0.08}$ & ${0.87}_{-0.08}^{+0.08}$ & ${0.89}_{-0.07}^{+0.07}$ & ${0.88}
_{-0.09}^{+0.08}$ \\
$\dot{s} \, (\rm{yr}^{-1})$ &  & ${-0.53}_{-0.24}^{+0.23}$ &  & ${-0.51}_{-0.27}^{+0.25}$ &  & ${-0.48}_{-0.25}^{+0.24}$ &  & ${-0.48}_{-0.25}^{+0.24}$ \\
$\dot{\alpha} \, (\rm{yr}^{-1})$ &  & ${3.9}_{-1.6}^{+0.9}$ &  & ${4.0}_{-1.5}^{+0.8}$ &  & ${-4.1}_{-0.7}^{+1.4}
$ &  & ${-4.0}_{-0.7}^{+1.4}$ \\
$\chi^2_{\rm min}$ & 6223.43 & 6210.66 & 6222.84 & 6210.78 & 6226.07 & 6214.22 & 6225.72 &  6213.40 \\
\hline
$\Delta \chi^2$                     & 0.6 & 0.0 & 0.0 & 0.1 & 3.2 & 3.6 & 2.9 & 2.7 \\
$\Delta \chi^2$ 2016 {\it Sp} & 0.1 & 0.0 & 0.0 & 0.0 & 4.0 & 3.3 & 3.5 & 2.8 \\
\hline
$M_1 \, ({\rm M_{\rm J}})$   &&  $15.5 \pm 1.6$ && $15.6 \pm 1.7$  && $15.8 \pm 1.4$ &&  $15.7 \pm 1.5$ \\
$M_2 \, ({\rm M_{\rm J}})$   &&  $11.8 \pm 0.7$ && $11.9 \pm 0.8$ && $12.0 \pm 0.6$ && $12.0 \pm 0.6$ \\
$D_L \, ({\rm kpc}$      &&  $3.03 \pm 0.19$ && $3.05 \pm 0.21$ && $3.09 \pm 0.16$ && $3.08 \pm 0.18$ \\
$r_\perp$ (AU)            &&  $0.42 \pm 0.03$ &&  $0.42 \pm 0.03$ && $0.43 \pm 0.03$ && $0.43 \pm 0.03$ \\
$\mu_{\rm hel,N}$ (mas yr$^{-1}$)  & & $-7.72$ & & $-7.64$ & & $8.04$  & & $8.07$ \\
$\mu_{\rm hel,E}$ (mas yr$^{-1}$)  & & $3.13$  & & $3.14$ & & $2.62$ & & $2.58$ \\
$\beta$ (N of E)    &&  $-61^{\circ}$  &&  $-61^{\circ}$ &&  $65^{\circ}$ &&    $65^{\circ}$  \\   
$\Delta \beta$ & & $-120^{\circ}$ & & $-120^{\circ}$ & & $5^{\circ}$ & & $6^{\circ}$ \\  
\hline
$P_{\rm lightcurve, rel}$ && 3.93 && 3.70 &&  0.664 && 1.0  \\
$P_{\rm Rich}$         &&  1.109 &&  1.109 &&  1.109 && 1.109  \\
$P_{\rm pm}$                  & & 0.00067 & & 0.00066 & & 0.473 & & 0.467 \\
$P_{\rm total, rel}$        & & 0.0057 & &  0.0052 & &  0.672 & & 1.0 \\
\hline
\enddata
\end{deluxetable}
\label{table:microlensing_results_green}

\newpage

\begin{deluxetable}{| l | cc | cc | cc | cc | }[h]
\rotate
\tabletypesize{\scriptsize}
\tablecaption{Microlensing parameters, physical parameters and relative probabilities  for the combined Spitzer and ground-based photometry for the B (yellow) solutions. 
The $+/-$ geometry refers
to the signs of $u_{0,Earth}$ / $u_{0,{\it Spitzer}}$. The left/right column for each geometry gives the parameters
for the models without/with lens orbital motion.}
\tablehead{
  & \multicolumn{2}{c|}{$+/+$} & \multicolumn{2}{c|}{$+/-$} & \multicolumn{2}{c|}{$-/+$} & \multicolumn{2}{c|}{$-/-$} 
}
\startdata
$\log_{\rm 10}s$ & ${-0.1856}_{-0.0023}^{+0.0025}$ & ${-0.1806}_{-0.0033}^{+0.0027}$ & ${-0.1857}_{-0.0021}^{+0.0023}$ & ${-0.1751}_{-0.0029}^{+0.0022}$ & ${-0.1855}_{-0.0022}^{+0.0022}$ & ${-0.1765}_{-0.0039}^{+0.0030}$ & ${-0.1858}_{-0.0023}^{+0.0023}$ & ${-0.1803}_{-0.0033}^{+0.0026}$ \\
$\log_{\rm 10}q$ & ${-0.150}_{-0.013}^{+0.012}$ & ${-0.119}_{-0.017}^{+0.015}$ & ${-0.153}_{-0.012}^{+0.012}$ & ${-0.124}_{
-0.016}^{+0.015}$ & ${-0.156}_{-0.012}^{+0.013}$ & ${-0.132}_{-0.014}^{+0.016}$ & ${-0.151}_{-0.013}^{+0.013}$ & ${-
0.119}_{-0.016}^{+0.016}$ \\
$log_{\rm 10}\rho$ & ${-1.601}_{-0.004}^{+0.004}$ & ${-1.570}_{-0.021}^{+0.014}$ & ${-1.598}_{-0.004}^{+0.004}$ & ${-1.529
}_{-0.019}^{+0.011}$ & ${-1.598}_{-0.004}^{+0.004}$ & ${-1.540}_{-0.023}^{+0.017}$ & ${-1.601}_{-0.004}^{+0.004}$ & 
${-1.569}_{-0.019}^{+0.013}$ \\
$u_{\rm 0}$ & ${1.174}_{-0.012}^{+0.011}$ & ${1.156}_{-0.012}^{+0.013}$ & ${1.191}_{-0.011}^{+0.010}$ & ${1.153}_{-0.009}^{+0.0
11}$ & ${-1.190}_{-0.010}^{+0.011}$ & ${-1.158}_{-0.015}^{+0.012}$ & ${-1.175}_{-0.011}^{+0.011}$ & ${-1.155}_{-0.013}
^{+0.011}$ \\
$\alpha$ & ${3.182}_{-0.014}^{+0.014}$ & ${3.143}_{-0.017}^{+0.019}$ & ${3.170}_{-0.013}^{+0.013}$ & ${3.134}_{-0.016}^{+
0.017}$ & ${3.107}_{-0.013}^{+0.013}$ & ${3.137}_{-0.015}^{+0.017}$ & ${3.101}_{-0.014}^{+0.014}$ & ${3.141}_{-0.019}^{+0
.018}$ \\
$t_{\rm 0}$ & ${7583.54}_{-0.05}^{+0.05}$ & ${7583.45}_{-0.07}^{+0.06}$ & ${7583.41}_{-0.06}^{+0.05}$ & ${7583.38}_{-0.06}^{+0.06}$ & ${7583.43}_{-0.05}^{+0.05}$ & $
{7583.40}_{-0.06}^{+0.06}$ & ${7583.53}_{-0.05}^{+0.06}$ & ${7583.45}_{-0.07}^{+0.06}$ \\
$t_{\rm E}$ & ${8.86}_{-0.06}^{+0.07}$ & ${8.75}_{-0.07}^{+0.09}$ & ${8.79}_{-0.06}^{+0.06}$ & ${8.57}_{-0.07}^{+0.08}$ & ${8.78}_{-0.07}^{+0.07}$ & ${8.58}_{-0.09}^{+0.09}$ & ${8.86}_{-0.07}^{+0.07}$ & ${8.74}_{-0.07}^{+0.09}$ \\
$\pi_{\rm E,E}$ & ${-0.043}_{-0.008}^{+0.008}$ & ${-0.053}_{-0.011}^{+0.013}$ & ${0.074}_{-0.024}^{+0.025}$ & ${0.05}_{-0.03}^{+0.03}$ & ${-0.056}_{-0.025}^{+0.026}$ & ${-0.07}_{-0.03}^{+0.03}$ & ${-0.032}_{-0.007}^{+0.008}$ & ${-0.040}_{-0.010}^{+0.011}$ \\
$\pi_{\rm E,N}$ & ${0.160}_{-0.006}^{+0.006}$ & ${0.171}_{-0.007}^{+0.007}$ & ${-1.857}_{-0.014}^{+0.014}$ & ${-1.810}_{-0.
019}^{+0.015}$ & ${1.847}_{-0.014}^{+0.014}$ & ${1.805}_{-0.019}^{+0.024}$ & ${-0.164}_{-0.006}^{+0.006}$ & ${-0.176}_{-0.007}^{+0.007}$ \\
$\dot{s} \, (\rm{yr}^{-1})$ &  & ${-0.59}_{-0.25}^{+0.25}$ &  & ${-0.41}_{-0.26}^{+0.24}$ &  & ${-0.32}_{-0.25}^{+0.25}$ &  & ${-0.59}_{-0.24}^{+0.25}$ \\
$\dot{\alpha} \, (\rm{yr}^{-1})$ &  & ${2.3}_{-1.6}^{+1.2}$ &  & ${5.8}_{-1.7}^{+1.0}$ &  & ${-4.8}_{-1.5}^{+2.0}$ &
  & ${-2.4}_{-1.1}^{+1.5}$ \\
$\chi^2_{\rm min}$ & 6240.84 & 6231.42 & 6232.70 & 6217.12 & 6230.30 & 6218.83  & 6241.06 & 6230.34 \\
\hline
$\Delta \chi^2$                     & 18.0 & 20.8 & 9.9 & 6.5 & 7.5 & 8.2 & 18.2 & 19.7 \\
$\Delta \chi^2$ 2016 {\it Sp} & 15.1 & 17.8 & 6.0 & 4.8 & 6.7 & 7.2 & 14.6 & 17.0 \\
\hline
$M_1 \, ({\rm M_{\rm J}})$   &&  $90 \pm 6$ && $8.1  \pm 0.4$  && $8.4 \pm 0.5$ &&  $89 \pm 6$ \\
$M_2 \, ({\rm M_{\rm J}})$   &&  $68 \pm 3$ && $6.09 \pm 0.19$ && $6.19 \pm 0.23$ && $68 \pm 3$ \\
$D_L \, ({\rm kpc}$      &&  $6.18 \pm 0.10$ && $2.06 \pm 0.07$ && $2.03 \pm 0.09$ && $6.18 \pm 0.10$ \\
$r_\perp$ (AU)            &&  $0.90 \pm 0.05$ &&  $0.276 \pm 0.016$ && $0.277 \pm 0.020$ && $0.89 \pm 0.05$ \\
$\mu_{\rm hel,N}$ (mas yr$^{-1}$)  & & $8.81$ & & $-8.39$ & & $8.85$  & & $-8.92$ \\
$\mu_{\rm hel,E}$ (mas yr$^{-1}$)  & & $-2.89$  & & $-1.80$ & & $-2.44$ & & $-2.23$ \\
$\beta$ (N of E)    &&  $107^{\circ}$  &&  $-88^{\circ}$ &&  $92^{\circ}$ &&    $-103^{\circ}$  \\   
$\Delta \beta$ & & $48^{\circ}$ & & $-148^{\circ}$ & & $33^{\circ}$ & & $-162^{\circ}$ \\  
\hline
$P_{\rm lightcurve, rel}$ && $1.22\times 10^{-4}$ && 0.156 &&  0.066 && $2.10\times 10^{-4}$  \\
$P_{\rm Rich}$         &&  1 &&  0.935 &&  0.935 && 1  \\
$P_{\rm pm}$                  & & 0.032 & &  0.0014 & & 0.123 & & $4.99\times 10^{-7}$  \\
$P_{\rm total, rel}$ & & $7.61 \times 10^{-6}$ & &  $3.93 \times 10^{-4}$ & &  $1.47 \times 10^{-2}$ & & $2.02 \times 10^{-10}$ \\
\hline
\enddata
\end{deluxetable}
\label{table:microlensing_results_yellow}

\end{document}